\documentclass[aps,pre,floats,twocolumn,showpacs,superscriptaddress]{revtex4-1}
\usepackage{amsmath}
\usepackage{graphicx}
\usepackage{amsfonts}
\usepackage[utf8]{inputenc}
\usepackage{cancel}
\usepackage{color}
\usepackage{hyperref}
\usepackage[]{algorithm2e} 
\usepackage{multirow}
\newcommand{\ave}[1]{\left \langle  #1 \right \rangle}
\newcommand{\p}[1]{\mathbf{#1}}

\begin{document}
\title{Role of adjacency matrix degeneracy in maximum-entropy-weighted network models}

\author{O. Sagarra}
\author{C. J. P\'erez Vicente}
\author{A.  D\'iaz-Guilera}
\affiliation{Departament de F\'{\i}sica Fonamental, Universitat de Barcelona, 08028 Barcelona, Spain}

\begin{abstract}
Complex network null models based on entropy maximization are becoming a powerful tool to characterize and analyze data from real systems. However, it is not easy to extract good and unbiased information from these models: A proper understanding of the nature of the underlying events represented in them is crucial. In this paper we emphasize this fact stressing how an accurate counting of configurations compatible with given constraints is fundamental to build good null models for the case of networks with integer valued adjacency matrices constructed from aggregation of one or multiple layers. We show how different assumptions about the elements from which the networks are built give rise to distinctively different statistics, even when considering the same observables to match those of real data. We illustrate our findings by applying the formalism to three datasets using an open-source software package accompanying the present work and demonstrate how such differences are clearly seen when measuring network observables.
\end{abstract}
\maketitle


\section{Introduction}
Network science~\cite{Newman2010a} is a prime example of the multiple uses that many tools and methodologies extracted from traditional physics can have when applied to a variety of transdisciplinar problems.

The advent of the so-called Big Data era given by the explosion of ICT technologies is providing researchers with unprecedented large datasets in a myriad of different fields ranging from biology~\cite{Serrano2012} to urban studies~\cite{sagarra2015a}, including bibliometrics~\cite{Menichetti2014a}, chemical sciences~\cite{guimera2005} or even history~\cite{schich2014a} to cite a few. The current challenge is to extract knowledge and useful information from such enormous datasets. A standard approach is based on representing them as a graph. Network representation of data is specially useful due to its relative simplicity in terms of computational effort, visualization~\cite{colomer2013deciphering} and analysis. However it presents some serious limitations which forces to look for innovative methodological tools.

One way to go beyond simple network representation is to generate appropriate null models. They must be flexible and reliable enough to compare our original data to in the search of {\em statistically relevant} patterns. In general, this is not a simple task, data processing is tricky and subtle in many situations and it may lead to wrong conclusions based on a poor understanding of the problem under study. A clever strategy to find efficient null models consists on generating randomized instances of a given network while keeping some quantities constant 
~\cite{colomer2013deciphering,Menichetti2014}. This can be done by algorithmic randomization of graphs~\cite{Molloy1995a} but such a procedure can be costly in terms of computational time (specially for dense datasets) and programming difficulty. Most importantly, most "rewiring techniques" do not always generate an unbiased sampling of networks~\cite{coolen2009constrained}.

A different approach to this problem has its roots in the analogy of networks with classical statistical mechanics systems~\cite{Park2004,bianconi2008entropies,Annibale2009a,roberts2011tailored,roberts2013tailored}, though it was originally proposed by sociologists and also by urban planners~\cite{wilson1970stp} under the name of {\em exponential random graphs}~\cite{holland1981}. It is based on the idea of constructing an ensemble of networks with different probabilities of appearance, which {\em on average} fulfil the considered constraints. The advantage of these methods is that they consider the possibility of having fluctuations (as usually happens in real data) and their derivation is completely analytical. Furthermore, such methods provide an easy way of rapidly simulating (and averaging) network instances belonging to a given ensemble. So far in the literature successful development of this kind of methodology has been performed for different types of monolayered networks~\cite{Annibale2009a,bianconi2009a,Garlaschelli2009b,sagarra2013smm}, directed \cite{roberts2011tailored} and bipartite structures~\cite{roberts2014}, stochastic block models~\cite{peixoto2012entropy} and some multiplex weighted networks~\cite{Menichetti2014a}.

Recently, there is a growing interest to study more complex mathematical structures~\cite{DeDomenico2013a,Cozzo2015} to account for the inherent multi-layered character of some network systems. This fact calls for the need of developing maximum entropy ensembles with a multi-layered perspective~\cite{Menichetti2014a}, which will help in the analysis of real world datasets. This is the main goal of this work. In this paper, we complement previous work on maximum entropy weighted networks by considering systems of aggregated multiplexes, where we have information about the layered structure of the system and the nature of their events, but -as usually happens for real data- we only have access to its accumulated structure (the sum of weights connecting two nodes in each layer for each pair of nodes). We show how the role of event and layer degeneracy induce important differences in the obtained statistics for each case, which recovers the mono-layered studied cases when the number of layers is set to unity. We further show that, despite the statistics being different, all the cases considered are examples of maximum likelihood networks of the dual problem~\cite{Garlaschelli2008a} but yield different expectations for network quantities, highlighting the need of {\em choosing an appropriate null model for each case study} based on the weighted character of the networks.

In section~\ref{sec:general} we present the mathematical framework and calculations of degeneracy for maximum entropy networks with arbitrary constraints. Section~\ref{sec:linear} extends the calculation to obtain explicit statistics for a very general case of constraints for the different cases considered. Finally, section~\ref{sec:data} presents an application of the model for the particular case of fixed strengths on the analysis of three real world datasets. Extended mathematical calculations are provided in the Appendix and details on the used datasets, measured quantities and numerical methods in the  Supplementary Material (SM) accompanying this work~\footnote{See Supplemental Material at [URL will be inserted by publisher] for details on the used datasets, measured network quantities and numerical methods used.}, including also a package for public use to apply the proposed models~\cite{ODME2014}.

\section{Maximum entropy constrained grand-canonical network ensembles with integer weights}\label{sec:general}
We consider a representation of a network of $N$ nodes, based on an adjacency matrix $\mathbf{T}$ composed by positive integer valued entries $t_{ij}\in \mathbb{N}$ which we call occupation numbers. Each of these entries accounts for the intensity of the interaction between any given pair of nodes $i$ and $j$ in the network, measured in terms of discrete events (which may be trips between locations in a mobility network or messages between users in social networks for instance). We study the case of directed networks with self-loops, albeit the undirected case follows from the derivation. Our objective is to fully determine the grand canonical ensemble of networks~\cite{Park2004,Bianconi2008a,sagarra2013smm} which fulfill {\em on average} some $Q+1$ given constraints $\{T \equiv \sum_{ij} t_{ij},C_q(\mathbf{T})\}$. The total number of events $T=\sum_{ij} t_{ij}$ determines the \textit{sampling} of the network, and is the minimal required constraint to consider any ensemble under the present framework~\cite{sagarra2013smm}. In this paper, we examine the problem where constraints take the form of linear combinations of functions of the individual occupation numbers $t_{ij}$,
\begin{equation}\label{eq_cons}
C_q(\mathbf{T}) =  \sum_{ij} c^{ij}_q f^{ij}_q(t_{ij}) \qquad \qquad c^{ij}_q \in \mathbb{R} \qquad \forall\, q\in Q.
\end{equation}
To {\em completely} determine an ensemble, it is not enough to specify the quantities we wish to fix (the constraints given by the original data), we must also define the statistical nature of the events allocated to the occupation numbers $t_{ij}$. In other words, we have to count all the network instances which give rise to the same particular configuration of the adjacency matrix $\mathbf{T}$. This {\em degeneracy term} $D(\mathbf{T})$ depends solely on the specifics of the system one represents, and counts the number of equivalent (micro) states that a particular unique realization of the adjacency matrix $\mathbf{T}$ can describe.

Once a grand canonical ensemble is fully constructed, the probability to obtain a particular configuration of occupation numbers $\mathbf{T}$ reads,
\begin{align}\label{eq_GC}
\begin{split}
P(\{\theta_q\},\mathbf{T}) &= \mathcal{Z}^{-1} D(\mathbf{T}) e^{H(\{\theta_q\},\p{T})} \\
H(\{\theta_q\},\p{T}) &\equiv \theta_T T(\mathbf{T}) +  \sum_q \theta_q C_q(\mathbf{T}).
\end{split}
\end{align}
The so-called Grand-Canonical partition function $\mathcal{Z} = \sum_{\{\mathbf{T}\}} D(\mathbf{T}) e^{H(\{\theta_q\},\p{T})}$ must be summed considering all the possible configurations of the adjacency matrix $\p{T}$ one can consider, regardless of whether the proposed constraints are met. Such a probability with an exponential form \cite{holland1981} is obtained by maximizing the Shannon entropy $S = \sum_{\{\mathbf{T}\}} P(\mathbf{T}) \ln P(\mathbf{T})$ associated to the ensemble while preserving the $Q+1$ constraints on average.

Using Equation \eqref{eq_cons} we reach,
\begin{align}\label{eq_GC_z}
\begin{split}
P(\mathbf{T}) &=  \mathcal{Z}^{-1} D(\mathbf{T}) \prod_{ij} z_{T}^{t_{ij}}z_{ij}(t_{ij}) \\
z_{ij}(t_{ij}) &\equiv \prod_q e^{\theta_q f_q^{ij}(t_{ij})} \qquad z_{T} \equiv e^{\theta}.
\end{split}
\end{align}
Let us notice that if the degeneracy term factorizes, i.e., $D(\mathbf{T}) \propto \prod_{ij} D_{ij} (t_{ij})$ the partition function can be re-expressed as
\begin{align}\label{eq_GC_ind}
\begin{split}
\mathcal{Z} &=  \prod_{ij} \mathcal{Z}_{ij} = \prod_{ij} \sum_{t_{ij} = 0}^{\infty} D_{ij}(t_{ij}) z_T^{t_{ij}}z_{ij}(t_{ij})  \\
 P(\mathbf{T}) &= \prod_{ij} p_{ij} (t_{ij})
=\prod_{ij} \frac{D_{ij}(t_{ij}) z_T^{t_{ij}}z_{ij}(t_{ij})}{\sum_{t'_{ij}=0}^{\infty}   D_{ij}(t'_{ij})z_T^{t'_{ij}} z_{ij}(t'_{ij}) },
\end{split}
\end{align}
where the statistics of $\mathbf{T}$ are formed by a set of independent random variables corresponding to the occupation numbers $\{t_{ij}\}$. Whenever one defines the degeneracy term and is able to sum the individual partition functions $\mathcal{Z}_{ij}$, then one gets the explicit statistics of the occupation numbers. The values of the Lagrange multipliers $(\theta_T,\{\theta_q\})$ associated to the $Q+1$ constraints (which can also be understood as {\em a posterior hidden variables}~\cite{Boguna2003,caldarelli2002a}) are obtained by solving the so called {\em saddle point} equations,
\begin{equation}\label{eq_saddle_general}
\begin{split}
\hat{C}_q &= \langle C(\mathbf{T}) \rangle = \sum_{ij}  \langle f^{ij}_q(t_{ij}) \rangle = \sum_{ij}  \sum_{t'_{ij}=0}^{\infty} p_{ij}(t'_{ij}) f^{ij}_q(t'_{ij})\\
\hat{T}  &= \sum_{ij} \ave{t_{ij}}.
\end{split}
\end{equation}
where $\{\hat{C}_q\}$ are the values of the quantities one wants to keep fixed (on average) in the ensemble.

The degeneracy terms are in general subtle to compute and to the best of our knowledge, are seldom considered in the literature. In order to calculate them, however, we need to make considerations about the type of networks at study and their elements. In this work, we consider systems composed by events that are either distinguishable or indistinguishable. Additionally, we study the general representation of an overlay multiplex network, which is obtained by aggregating several layers of a system into a single (integer weighted) adjacency matrix~\cite{DeDomenico2013a,Cozzo2015}. Examples of such networks range from aggregation of transportation layers~\cite{Cardillo2013}, networks generated by accumulation of information over a certain time span such as Origin-Destination matrices~\cite{sagarra2015a}, email communications~\cite{guimera2002a} or human contacts~\cite{gauvin2013a} and even an aggregation of trading activities in different sectors such as the World Trade Network~\cite{Squartini2011a}.

Thus the system under consideration is an aggregation of $M$ network layers containing the same type of events: They can be either a group of layers composed by distinguishable (Multi-Edge - ME) or indistinguishable (Weighted - W) events or even an aggregation of Binary (B) networks. The occupation numbers corresponding to layer $m$ are noted $t_{ij}^m$, but we only have {\em access} to information about their accumulated value through all the layers, i.e. the aggregated occupation numbers $t_{ij} =  \sum_m t_{ij}^m$.

Finally, the degeneracy term is the product of the multiplicity induced by the nature of the events times the nature of the layers (which in the only real possible scenario are always distinguishable) $D(\mathbf{T}) = D(\mathbf{T})_{Events}D(\mathbf{T})_{Layers}$. This last term is computed (for each pair of nodes or state $ij$) by counting the number of different groupings one can construct by splitting $t_{ij} = \sum_{m} t^m_{ij}$ (distinguishable or indistinguishable) aggregated events into $M$ different layers respecting the occupation limitation of the considered events: Either only one event per layer (Binary network) or an unrestricted number (Weighted and Multi-Edge networks). 
\begin{table}[htbp]
\begin{center}
\scalebox{0.9}{
\begin{tabular}{l | l |l }
Network Type &  $D(\mathbf{T})_{Events}$ & $D(\mathbf{T})_{Layers}$ \\\hline
Multi-Edge (ME)& $\frac{T!}{\prod_{ij} (\sum_m t_{ij}^m)!}$&$\prod_{ij}\sum_{\{t_{ij}^m\}} \frac{\left (\sum_m t_{ij}^m \right )!}{\prod_{m} t_{ij}^m!} = \prod_{ij} M^{\sum_m t^m_{ij}}$\\
Weighted (W) &1 & $\prod_{ij} \binom{M+\sum_m t_{ij}^m -1 }{\sum_m t_{ij}^m}$\\
Binary Dist. (BD)&$T!$ & $\prod_{ij} \binom{M}{\sum_m t_{ij}^m}$\\
Binary Indist. (BI)&1 & $\prod_{ij} \binom{M}{\sum_m t_{ij}^m}$\\
\end{tabular}
}
\caption{degeneracy terms corresponding to the elements of the system and their layers for each case.}
\label{table_deg_terms}
\end{center}
\end{table}

The resulting degeneracy terms are shown in table \ref{table_deg_terms} (see details in Appendix~\ref{app_4}), from which one can see that in principle the event degeneracy term does not factorize for the distinguishable cases due to the presence of the variable $T! = \left(\sum_{ij} t_{ij} \right)!$. 
One can nevertheless obtain an \emph{effective} degeneracy term by substituting it by $\hat{T}!$ (a constant) -as shown in Appendix~\ref{app_4}, where a complete discussion of the implications of this substitution for the different cases is provided- which leads to results fully equivalent to those obtained by performing the exact calculation for the Multi-Edge case with constraints of the form~\eqref{eq_fij_linear}. In doing so, 
two preliminary conclusions can be drawn. Firstly, both the distinguishable and indistinguishable binary cases will lead to the same statistics since their degeneracy term on events will be constant (hence on the remainder of the paper we will omit the case BD). Secondly, in all the cases the complete degeneracy terms will factorize into state $ij$ independent terms, which means that the statistics of the aggregated occupation numbers will be state independent (equation \eqref{eq_GC_ind}).

\section{Linear constraints on aggregated occupation numbers}\label{sec:linear}

To go further in our derivation, we now consider the case where the constraints are linear functions of the aggregated occupation numbers,
\begin{align}\label{eq_fij_linear}
f^{ij}_q(t_{ij}) = c_q^{ij}t_{ij} = c_q^{ij}\sum_m t_{ij}^m.
\end{align}
Such a case is very generic and includes networks with local constraints on nodes~\cite{Squartini2011c}, community constraints~\cite{bianconi2009a} and generalized cost constraints such as distances~\cite{halu2014a}. The case where the constraints depend on both the binary projection of the occupation numbers and their values $f_{ij}(t_{ij}) = c^{ij}_q t_{ij} + \tilde{c}^{ij}_q \Theta(t_{ij})$ can be derived from the methodology developed here and is analyzed in Appendix \ref{app_1}.

The individual partition functions can be summed
\begin{align}\label{eq:part_funcs}
\begin{split}
&\mathcal{Z}_{ij} = \sum_{t_{ij}} D_{ij} (t_{ij})  z_{ij}^{t_{ij}} = \\
&=\left \{ \begin{array}{l l l}
\text{ME:}&\sum_{t_{ij}=0}^{\infty} \frac{(Mz_{ij})^{t_{ij}}}{t_{ij}!} = e^{Mz_{ij}}&\\
\text{W:}&\sum_{t_{ij} = 0}^{\infty} \binom{M+t_{ij} -1 }{t_{ij}} z_{ij}^{t_{ij}} = (1-z_{ij})^{-M};& z_{ij}<1 \\
\text{B:}& \sum_{t_{ij} = 0}^M \binom{M}{t_{ij}}  z_{ij}^{t_{ij}} = (1+z_{ij})^M;& t_{ij}\leq M
\end{array}
\right . .
\end{split}
\end{align}
In this case, we have redefined $z_{ij} \equiv e^{\theta_T} \prod_{q}^Qe^{\theta_q c_q^{ij}}$ to ease notation. This leads to,
\begin{align}\label{eq_all_stats}
\begin{split}
p_{ij}^{ME}(t_{ij}) &= e^{-Mz_{ij}} \frac{(Mz_{ij})^{t_{ij}}}{t_{ij}!}\\
p_{ij}^{W}(t_{ij}) &= \binom{M+t_{ij} -1 }{t_{ij}} z_{ij}^{t_{ij}} (1-z_{ij})^M\\
p_{ij}^{B}(t_{ij}) &= \binom{M}{t_{ij}} \left (\frac{z_{ij}}{1+z_{ij}} \right)^{t_{ij}} (1+z_{ij})^{-(M-t_{ij})}.
\end{split}
\end{align}
And we recover well known probability distributions: Poisson distribution for the Multi-Edge case~\cite{sagarra2013smm} (independent of the number of layers $M$), Negative Binomial for the Weighted case (being the geometric distribution~\cite{Garlaschelli2009b} a special case when $M=1$) and Binomial distribution for the aggregated Binary case (being the Bernoulli distribution~\cite{Park2004} a special case for $M=1$).

The resulting statistics show some important features: On the one hand, one sees that albeit the degeneracy term changes for Multi-Edge networks for either case of a monolayer or a multilayer, the form of the obtained statistics does not. This means that {\em it is not possible to distinguish a Multi-Edge monolayered network from an aggregation of multiple Multi-Edge layers belonging to an ensemble with the same constraints}. On the other hand, the situation for the other cases changes: For multiplexes the resulting occupation numbers will have different statistics from the monoplex case. This has the implication than one could in principle {\em discern} the aggregated nature of a network by inspection of their accumulated edge statistics $\{t_{ij}\}$, provided that one has access to enough realizations of a system and that it belongs to the same ensemble (i.e. the system evolves according to some given, even if unknown, linear constraints~\cite{sagarra2015a} of the form in equation \eqref{eq_fij_linear}).

\begin{table}[htbp]
\begin{center}
\begin{tabular}{l |l l l | l}
Network Type & $\ave{t_{ij}}$ & $\sigma^2_{t_{ij}}$ & $\frac{\sigma^2_{t_{ij}}}{\ave{t_{i}}^2}$&Domain $z_{ij}$\\ \hline
ME&$Mz_{ij}$&$Mz_{ij}$&$(Mz_{ij})^{-1}$&$[0,\infty)$\\
W&$M\frac{z_{ij}}{1-z_{ij}}$&$M\frac{z_{ij}}{(1-z_{ij})^2}$&$(Mz_{ij})^{-1}$ &$[0,1)$\\
B&$M\frac{z_{ij}}{1+z_{ij}}$&$M\frac{z_{ij}}{(1+z_{ij})^2}$&$(Mz_{ij})^{-1}$&$[0,\infty)$
\end{tabular}
\caption{First and second moment of the considered distributions, together with the relative fluctuations.}
\label{table_2moments}
\end{center}
\end{table}

Another important implication of the obtained statistics is the very different interpretations encoded in the values $z_{ij}$. This collection of values is related to the constraints originally imposed to the network ensemble through the set of Lagrange multipliers $(\theta_T,\{\theta_q\})$ (equations \eqref{eq_GC_z} and \eqref{eq_saddle_general}) and can be understood as \emph{a posteriori} measures related to the intensity of each node-pair $ij$. These measures encode the correlations between nodes imposed by the constrained topology (note that for local constraints only at the level of nodes we obtain a factorization $\ave{t_{ij}}=M x_i y_j$).
Table \ref{table_2moments} reports the two first central moments of each distribution. For the Multi-Edge case $z_{ij}$ is both directly mapped to the average occupation of the considered link $ij$, $\ave{t_{ij}}$ and to its (relative) importance in the network~\cite{sagarra2013smm}. In all the other cases, however, $z_{ij}$ relates to a probability of a set of events emerging from a given node, to be allocated to a link $ij$. Obviously, as $\ave{t_{ij}}$ grows, $z_{ij}$ grows in all cases, but not in the same linear way (in the W case, for instance, $z_{ij}$ is bounded to a maximum value of 1). This means that while in all cases $z_{ij}$ is related to the importance of a given link with respect to the others, the dependency in all non ME cases is highly non-linear. Finally, we can see that for a large number of layers $M \gg T/N^2$, the ensembles become equivalent to the ME, as the degeneracy term on (distinguishable) layers dominates the configuration space of the ensembles.

The different obtained statistics are highly relevant as their marked differences point out at a 
(regularly overlooked) problem: Different maximum entropy ensembles yield very different statistics for the same considered constraints, and hence \textit{each dataset needs to be analyzed carefully, since the process behind the formation of each network dictates their degeneracy term}. Furthermore, all the obtained statistics are derived from a maximum entropy methodology, and hence the values $z_{ij}$ obtained from \eqref{eq_saddle_general} are in all cases maximum likelihood estimates for the probability of $\p{T}$ to belong to the set of models described by equation \eqref{eq_all_stats} (see Appendix \ref{app_2}). Thus, any of the presented models will be a {\em correct} ensemble in a maximum likelihood sense~\cite{Garlaschelli2008a} for some given constraints, and the appropriate choice for each network representation depends on the system under study, in contrast to the interpretation given by~\cite{Squartini2011c}.

This means that if one wants to assess the effects a given constraint has on a network constructed from real data, \textit{one needs to very carefully choose the appropriate null model to compare the data to}.
It is also worth pointing out that most of these ensembles are not equivalent to a manual rewiring of the network~\cite{sagarra2014efn} (albeit one expects small differences, see Appendix \ref{app_3}). However, maximum entropy models allow for an analytical treatment of the problem, and simplify the generation of network samples when the considered constraints are increasingly complicated (both at the coding level and at the computational one). This has many implications, including the possibility of computing $p$ values, information theoretic related quantities such as ensemble entropies~\cite{bianconi2009a,Annibale2009a,roberts2011tailored,roberts2013tailored,peixoto2012entropy} or model likelihoods as well as efficient weighted network pruning algorithms~\cite{Dianati2015a,Serrano2009}. Moreover, this procedure helps in the fast and simple generation of samples of networks with prescribed constraints.

The main difficulty of the \emph{soft-constrained} maximum entropy framework hereby presented for null model generation is the problem of solving the saddle point equations \eqref{eq_saddle_general} associated to each ensemble. With the exception of some particular cases~\cite{sagarra2014efn}, these equations do not have an analytical solution and must be obtained numerically. In such a case, the best approach is to maximize the associated loglikelihood of each model to a set of observations (constraints), yet the difficulty of each problem increases with the number of constraints since each fixed quantity has an associated variable to be solved. Considering the different statistics obtained in this paper, the most difficult case by far is the Weighted one (W), since the condition that $0\leq z_{ij} <1$ imposes a non-convex condition in the domain of the loglikelihood function to maximize, while the others are in general easily solved using iterative balancing algorithms (see SM for an extended discussion and details on the numerical methods used).

\section{Application to real data: The case of fixed strengths}\label{sec:data}
To highlight the importance of considering an appropriate null model for the assessment of real data features, in this final section we consider the case of networks with a fixed strength sequence. Real networks usually display highly skewed node strength distributions, which have important effects in their observables. Hence, to correctly assess whether some observed feature in a dataset can be solely explained by the strength distribution, it is crucial to choose an appropriate null model to compare the data to. This situation is especially important for instance with regard to community analysis through modularity maximization for weighted networks, because the modularity function to be optimized~\cite{bianconi2009a} needs as input a prediction from a null model with fixed strengths ($Q \propto \sum_{i,j} \left(\hat{t}_{ij} - \ave{t_{ij}}\right)\delta_{c_i,c_j}$ where $\{c\}$ are the community node labels associated to the optimal network partition).
For a directed model with fixed strengths, the constraints in equation \eqref{eq_cons} read ($\theta_T$ is not needed because $T=\sum_i s_i^{in} = \sum_j s_j^{out}$),
\begin{align}
\begin{split}
s^{out}_q &= \sum_{ij} c^{ij}_q t_{ij} = \sum_{ij} \delta_{qi} t_{ij} = \sum_j t_{qj} \qquad \forall q =1..N\\
s^{in}_r  &= \sum_{ij} c^{ij}_r t_{ij} = \sum_{ij} \delta_{jr} t_{ij} = \sum_i t_{ir} \qquad \forall r =1..N\\
z_{ij} &= \prod_{q}e^{\theta_q \delta_{iq} } \prod_{r}e^{\theta'_r \delta_{rj}} = x_i y_j \qquad x_q \equiv e^{\theta_q} \, , \, y_r \equiv e^{\theta'_r}.
\end{split}
\end{align}
So the resulting saddle point equations \eqref{eq_saddle_general} are
\begin{align}
\hat{s}^{out}_i = \ave{s^{out}} \qquad \hat{s}^{in}_i = \ave{s^{in}} \qquad i=1...N
\end{align}
where $\hat{s}$ denotes the numerical value found in the data for the random variable $s$, which particularized to each case reads,
\begin{align}
\begin{array}{l l}
\text{ME:} & \left\{ \begin{array}{l}
\hat{s}^{out}_i = Mx_i \sum_{j} y_j \\
\hat{s}^{in}_j = My_j \sum_{i} x_i
\end{array}\right.
 \\
\text{W:} & \left\{ \begin{array}{l}
\hat{s}^{out}_i = Mx_i \sum_j \frac{y_j}{1-x_iy_j} \\
\hat{s}^{in}_j  = My_j \sum_i \frac{x_i}{1-x_iy_j}
\end{array} \right .
\\
\text{B:}&  \left\{ \begin{array}{l}
\hat{s}^{out}_i = Mx_i \sum_j \frac{y_j}{1+x_iy_j} \\
\hat{s}^{in}_j  = My_j \sum_i \frac{x_i}{1+x_iy_j}
\end{array} \right.
\end{array} .
\end{align}
The ME case has an analytical solution~\cite{sagarra2014efn} while the others must be solved computationally. Supplementary Material SM provides extended details about the network quantities computed, simulations, averaging and computational methods and algorithms used in this section, which are available in the freely provided, open source ODME package~\cite{ODME2014}).

\begin{figure*}[htbp]
\begin{center}
\includegraphics[width=0.8\textwidth]{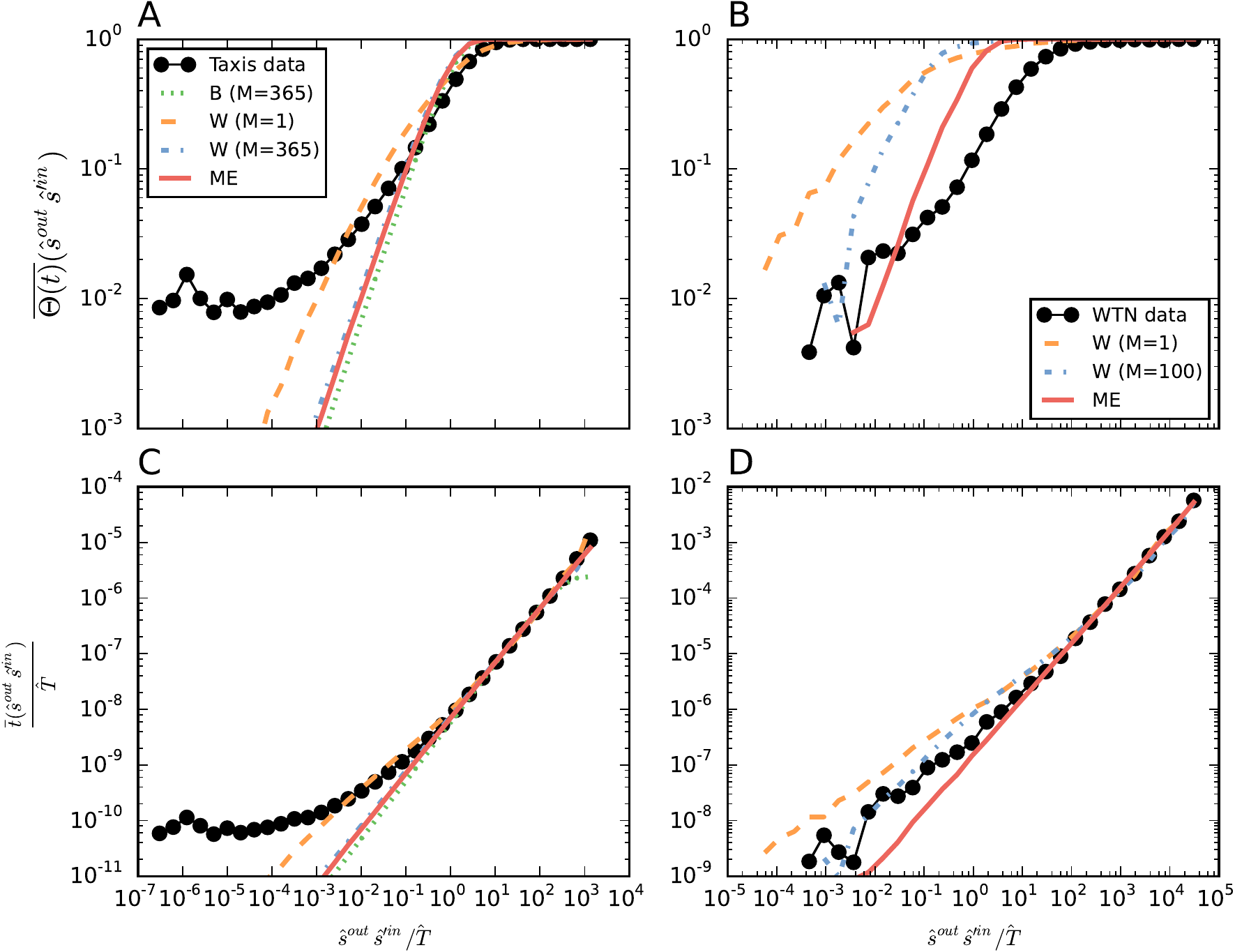}
\caption{(Color online) \textbf{Node pair statistics:} Binary connection probability (top) and rescaled average edge weight (bottom) as function of product of origin and destination node strength.  Results averaged over $r=5\cdot 10^2$ and $r=10^4$ realizations for the different models respectively with applied log-binning. The sudden increase for the binary pair-node connection probability can be clearly seen for the W case.}
\label{fig:edges2}
\end{center}
\end{figure*}
As real world datasets we use a snapshot of the World Trade Network (WTN), the OD matrix generated by Taxi trips in Manhattan for the year 2011~\cite{Santi14,sagarra2015a} and the multiplex European Airline Transportation network~\cite{Cardillo2013}. WTN has been vastly studied in the literature and recently has been represented as an aggregated system of $M \sim 100$ layers~\cite{Mastrandrea2014a} representing different types of commodities being traded. In this network, nodes represent countries and weights represent the amount of trade between them, measured in millions of US dollars. In the OD Taxi dataset, which we construct as the aggregation of $M=365$ daily layer snapshots, each node represents an intersection and each weight the number of trips recorded between them~\cite{sagarra2011sca}. Finally, in the airline network each node is an airport and weights correspond to the number of airlines providing direct connections between them, so the network is an aggregation of $M=37$ binary layers (one for each airline).

In all cases we will consider directed networks, and throughout this paper we will only show results in the outgoing direction, as the results in the incoming direction are qualitatively equal. Note that the binary aggregated case cannot be always applied since the maximum number of events allocated per node pair cannot exceed the number of layers, and for the WTN dataset this condition ($\max(\{\hat{s}_i\}) \leq s^{max} = NM$) is violated for some nodes. 

To analyze the difference between models, we compute ensemble expectations for different edge and node related properties suitably rescaled (fixing the original strength distribution of each dataset) and then compare the obtained results with the real observed data features. 

The airline dataset is very sparse and differences between models are not wide, which points out the need of adequate sampling for a successful analysis on weighted networks. Anyhow, since it is by construction an aggregation of binary layers, the B case displays the most resemblance to the data, both qualitatively and quantitatively (see SM).

\begin{figure*}[htbp]
\begin{center}
\includegraphics[width=0.8\textwidth]{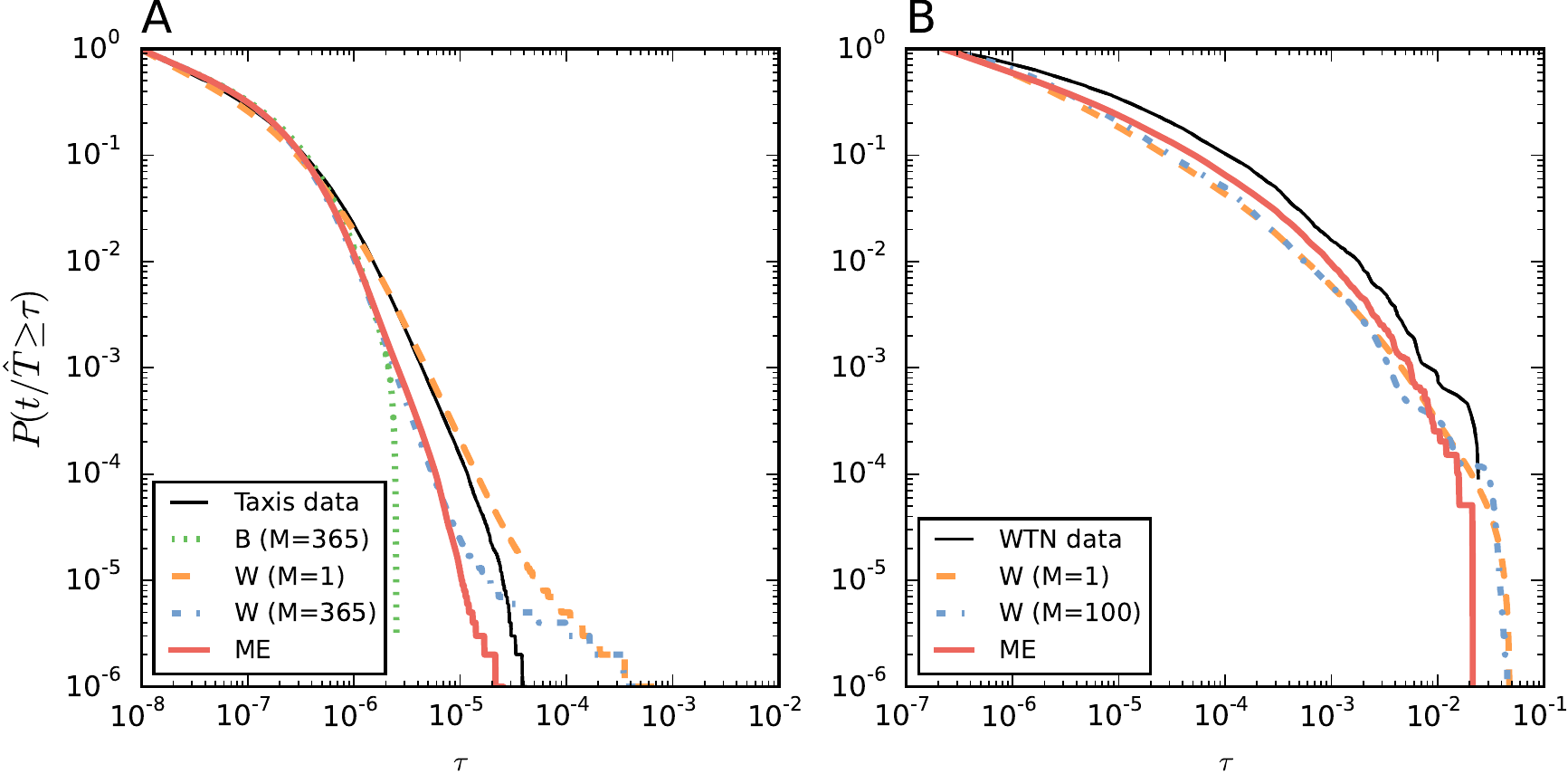}
\caption{(Color online) \textbf{Weight statistics:} Existing node pair weight complementary cumulative distribution for the Taxi (left) and WTN (right) datasets. Same conditions as Figure~\ref{fig:edges2} apply. The presence of extremely large weights can be seen in the tails of the distributions for both the W monolayer and multilayer case.}
\label{fig:edges}
\end{center}
\end{figure*}

\begin{figure*}[htbp]
\begin{center}
\includegraphics[width=0.8\textwidth]{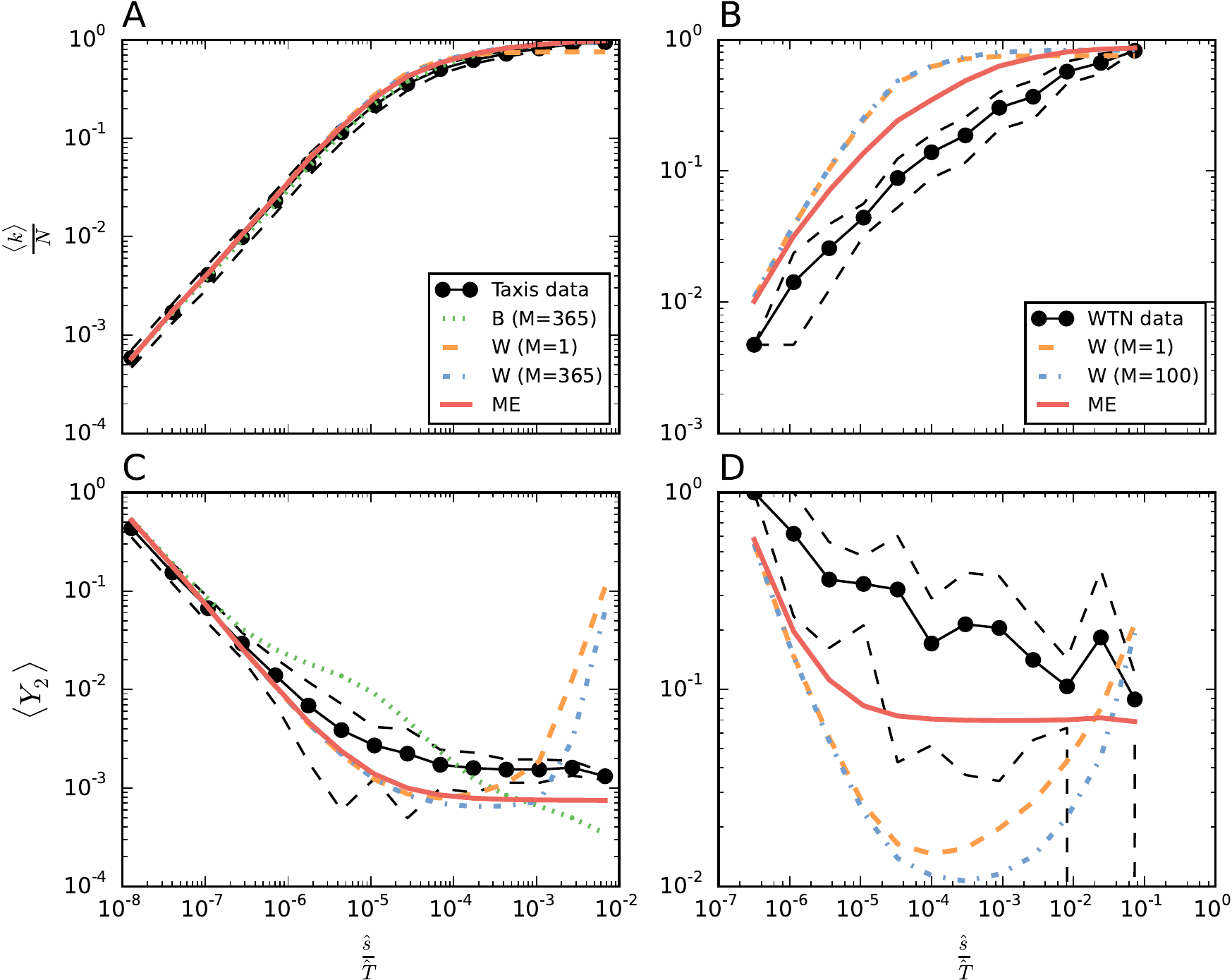}
\caption{(Color online) \textbf{First order node statistics:} Rescaled degree (top) and disparity (bottom) for the Taxi (left) and WTN (right) datasets. Same conditions as Figure~\ref{fig:edges2} apply. Dashed lines represent log-binned standard deviation ranges for the real data. The U-shaped disparity profile is clearly seen for the W cases in sharp contrast with the monotonous behaviour of both the real data and the ME model.}
\label{fig:nodes}
\end{center}
\end{figure*}

The WTN and Taxi datasets, in contrast, contain enough sampling for the wide differences between models to emerge. All cases have the same number of events $\hat{T}$ on average, but they are not distributed among connections between nodes in the same way for the different models. Being zero the most probable value for the geometric distribution, for the W case with a single layer the connection probability initially grows distinctively faster than in all the other cases leading to larger number of binary connections between low strength nodes (Figure~\ref{fig:edges2}-A,B). Yet the higher relative fluctuations of the geometric statistics also generate extremely large maximum weights in the tail of the existing occupation number distribution (Figure~\ref{fig:edges}), which are concentrated in connections between high strength nodes (Figures~\ref{fig:edges2}-C,D). Since the total number of events incoming and outgoing each node is fixed, this means that the W case has comparatively the lowest degrees for the most weighted nodes despite counting the larger number of binary connections $E = \sum_{ij} \Theta(t_{ij}) = \sum_i k^{out}_i = \sum_j k^{in}_i$ as can be seen in Figure~\ref{fig:nodes}-A,B.

\begin{figure*}[htbp]
\begin{center}
\includegraphics[width=0.8\textwidth]{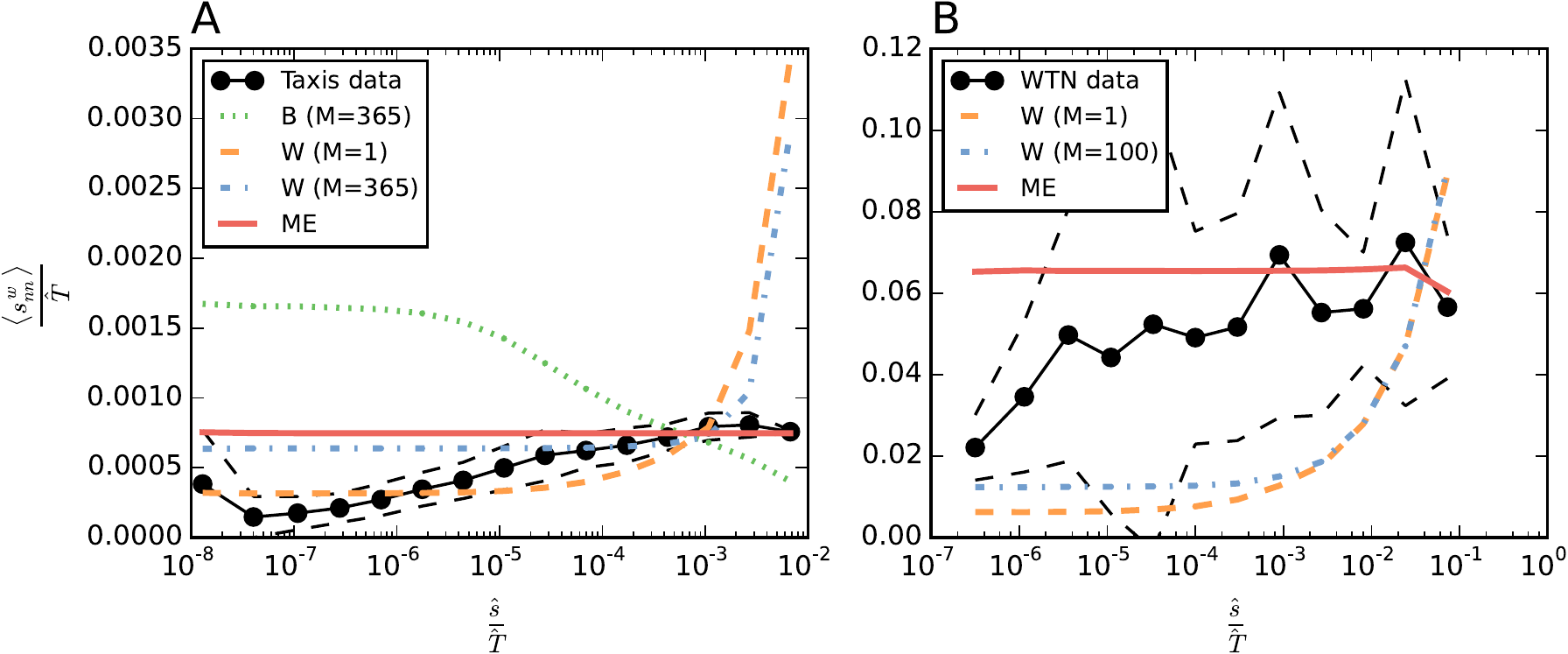}
\caption{(Color online) \textbf{Second order node statistics:} Rescaled weighted average strength (bottom) for the Taxi (left) and WTN (right) datasets. Same conditions as Figure~\ref{fig:edges2} apply. Dashed lines represent log-binned standard deviation ranges for the real data. A sharp increase is clearly seen for high strength nodes in the W cases.}
\label{fig:nodes2}
\end{center}
\end{figure*}

These anomalies for low and high strength nodes respectively for the W case produce wild asymmetries in the allocation of weights per node, which can be studied measuring their disparity $Y_2 = \sum_{j} t^2_{ij}/\left(\sum_j t_{ij}\right)^2$ (Figure \ref{fig:nodes}-C,D), which quantifies how homogeneously distributed are the weights emerging from each node: It displays a U shaped form with both low and high strength nodes tending to very strongly concentrate their weights on few connections. This non-monotonic behaviour is in strong contrast with the one observed for the real data and usually in other datasets~\cite{Menichetti2014a}. Concerning second order node correlations, the outgoing weighted average neighbor strength $s^w_{nn} = \sum_j t_{ij} s^{in}_j/s^{out}_i$ (Figure~\ref{fig:nodes2}) again displays a large range of variation for the W case (with either one or more than one layer) in contrast with the slight assortative profile of the real data, the uncorrelated profile of the ME case and the slight dissassortative trend of the B case. This last case is caused by the combination of two factors: The limitation on the maximum weight of the edges cannot compensate (with large weights connecting the nodes with the larger strength) the tendency of large nodes to be connected to a macroscopic fraction of the network, which is dominated by low strength nodes.

Obviously none of the null models used reproduce the real data, however, the goal in model construction is rather to assess the structural impact that a given constraint (in this case a strength distribution) has on the network observables. In this sense, we show that different models provide very different insights about such impacts. In particular, since the Airline dataset is by construction an aggregated binary network and the Taxi dataset a Multi-Edge one (people riding Taxis are clearly distinguishable), the fact that the B and ME cases respectively lie closer to the real data comes at no surprise. The WTN case, however, is unclear: The modelling of trade transactions has not a clearly defined nature, but if one assumes the WTN to be a multilayered network, its aggregated analysis should be performed using either the W (with $M>1$ layers) or ME case, which again are closer both in functional form and qualitative values to the real case (in contrast with~\cite{Mastrandrea2014a} where the W model with a single layer is used).

\section{Conclusions}

In this work we have showed the importance of considering the nature of the events one wishes to model when using an integer valued network representation of a system. We have developed and solved a maximum entropy framework for model generation applied to networks generated by aggregation~\cite{DeDomenico2013a,Cozzo2015} of multiplexes. We have shown how different considerations about the nature of the events generating the elements of the multiplex give rise to distinctively different node pair statistics. For the case where one wants to fix properties expressed as linear functions of the individual weights (and optionally their binary projection) in the network, we elegantly recovered well know statistics such as Poisson, Binomial, Negative Binomial, Geometric and Bernouilli for each case.

We have further provided a practical example by focusing on the case of fixed strengths and applying the models to assess relevant features on three different real-world datasets containing different types of weights, showing how the role of adjacency matrix degeneracy plays a crucial role in model construction. To this end, we have made considerations about the statistical nature of the obtained models as well as the weaknesses and strengths derived from their practical applications in real cases. Finally, we provide the open source software package ODME~\cite{ODME2014} for practitioners to apply the proposed models to other datasets.

The insights derived from this paper can open the door to the objective identification of truly multiplex structures (except one case where it has been shown to be impossible) by inspection of the statistics of their edges, provided that several instances of a network belonging to the same ensemble are available.

The take home message of this work is that in order to perform a meaningful analysis on a given network, a practitioner needs to be able to select an appropriate null model, which not only depends on the endogenous constraints one considers but also on the very nature of the process one is modelling. Our work provides researchers with a range of maximum entropy (and maximum likelihood) models to choose from, covering a wide spectra of possibilities for the case of weighted networks. Each of this models is not {\em wrong or even right} in a general case despite yielding very different predictions for the same sets of constraints, but just more or less appropriate depending on the problem at hand one wants to study.

\section*{Acknowledgements}
We thank A. Allard, A. Fern\'andez, F. Font-Clos and R. Toral for useful comments and suggestions. This work has been partially supported by the LASAGNE (Contract No.318132) and MULTIPLEX (Contract No.317532) EU projects, the Spanish MINECO (FIS2012-38266) and by the Generalitat de Catalunya (2014SGR608). O.S. acknowledges financial support from Generalitat de Catalunya (FI-program) and the Spanish MINECO (FPU-program).

\appendix

\section{Binary constraints}\label{app_1}
We develop hereby the general case where the constraints have the general form $f^{ij}_q(t_{ij}) = \tilde{c}^{ij}_{q} \Theta(t_{ij}) + c^{ij}_q t_{ij}$. The general derivation remains essentially unchanged from the linear case (main text) with a slight modification in the calculation of the explicit partition function,
\begin{align}
\begin{split}
\mathcal{Z}_{ij} &= \sum_{t_{ij}=0}^\infty D_{ij}(t_{ij}) \tilde{z}_{ij}^{\Theta(t_{ij})} z_{ij}^{t_{ij}} =\\
&= \tilde{z}_{ij}  \left( \sum_{t_{ij}=0}^{\infty} D_{ij}(t_{ij}) z_{ij}^{t_{ij}} - D_{ij}(0) \right) + D_{ij}(0)\\
z_{ij} &\equiv e^{\theta_T} \prod_q e^{\theta_q c_q^{ij}} \qquad \tilde{z}_{ij} \equiv \prod_q e^{\theta_q  \tilde{c}_q^{ij}}.
\end{split}
\end{align}
Where $\{z_{ij}\}$ have been redefined and the constraint on the total number of events $T=\sum_{ij} t_{ij}$ is introduced in their redefinition. This yields,
\begin{align}
p_{ij}(t_{ij}) = \frac{D_{ij}(t_{ij})\tilde{z}_{ij}^{\Theta(t_{ij})}z_{ij}^{t_{ij}}}{\tilde{z}_{ij}  \left( \sum_{t'_{ij}=0}^{\infty} D_{ij}(t'_{ij}) z_{ij}^{t'_{ij}} - D_{ij}(0) \right) +D_{ij}(0) }
\end{align}
which corresponds to the zero-inflated versions (ZI) of the previous statistics recovered in the case $f_{ij} = c^{ij}_q t_{ij}$, that is, asymmetric statistics where the probability of the first occurrence is different from the rest. Note that in this very general case, one can always set $\{c^{ij}_q=0\,\forall ij\}$ to include the case where only binary constraints are considered). Explicitly, for the different statistics, we have,
\begin{align}
\begin{split}
&\text{ME (ZIP - Zero Inflated Poisson):}\\
&p_{ij}(t_{ij}) = \frac{(Mz_{ij})^{t_{ij}}}{t_{ij}!} \frac{\tilde{z}_{ij}^{\Theta(t_{ij})}}{\tilde{z}_{ij} \left( e^{Mz_{ij}} -1 \right)+1} \\
&\text{W (ZINB - Zero Inflated Negative Binomial):} \\
&p_{ij}(t_{ij}) = \binom{M+t_{ij} -1 }{t_{ij}}z_{ij}^{t_{ij}} \frac{\tilde{z}_{ij}^{\Theta(t_{ij})}}{\tilde{z}_{ij} \left( (1-z_{ij})^{-M}-1 \right)+1}\\
&\text{B (ZIB - Zero Inflated Binomial):}\\
&p_{ij}(t_{ij}) = \binom{M}{t_{ij}} z_{ij}^{t_{ij}} \frac{\tilde{z}_{ij}^{\Theta(t_{ij})}}{\tilde{z}_{ij} \left( (1+z_{ij})^M -1 \right) +1} .
\end{split}
\end{align}

We can then compute the binary connection statistics,
\begin{align}
\begin{split}
\ave{\Theta(t_{ij})} = 1- p_{ij}(0) =  \left \{ \begin{array}{l l}
\text{ME}& \frac{\tilde{z}_{ij} (e^{Mz_{ij}} -1)}{\tilde{z}_{ij} (e^{Mz_{ij}} -1)+1}\\
\text{W}& \frac{\tilde{z}_{ij} ((1-z_{ij})^{-M}-1)}{\tilde{z}_{ij} ((1-z_{ij})^{-M}-1)+1} \\
\text{B}& \frac{\tilde{z}_{ij} \left( (1+z_{ij})^M -1 \right)}{\tilde{z}_{ij} \left( (1+z_{ij})^M -1 \right)+1}. \\
\end{array} \right.
\end{split}
\end{align}

Note how the binary projection in all cases corresponds to Bernouilli statistics. Regarding the occupation number statistics, one has explicitly,
\begin{align}
\begin{split}
\ave{t^+_{ij}} &\equiv \ave{t_{ij} | t_{ij}>0} = \\
&=\left \{ \begin{array}{l l}
\text{ME}& M\frac{z_{ij}}{1-e^{-Mz_{ij}}} \\
\text{W}& M \frac{z_{ij}}{1-z_{ij}} \frac{1}{\left( 1 - (1-z_{ij})^{M}\right)}\\
\text{B}& M \frac{z_{ij}}{1+z_{ij}} \frac{1}{\left( 1 - (1+z_{ij})^{ -M}\right)}\\
\end{array} \right. \\
\ave{t_{ij}} &= \ave{\Theta(t_{ij})} \ave{t_{ij}^+} = \\
&=\left \{ \begin{array}{l l}
\text{ME}& Mz_{ij}^{a^{ij}_q+\tilde{a}^{ij}_{q}} \frac{e^{Mz_{ij}}}{1+\tilde{z}_{ij} (e^{Mz_{ij}}-1)}\\
\text{W}& M \frac{z_{ij}^{a^{ij}_q+\tilde{a}^{ij}_{q}}}{1-z_{ij}} \frac{1}{(1-z_{ij})^M + \tilde{z}_{ij}(1- (1-z_{ij})^M)}\\
\text{B}& M \frac{z_{ij}^{a^{ij}_q+\tilde{a}^{ij}_{q}}}{1+z_{ij}} \frac{1}{(1+z_{ij})^{-M} + \tilde{z}_{ij} \left( 1-(1+z_{ij})^{-M} \right)} .
\end{array} \right. \\
\end{split}
\end{align}

And we observe a clear non-trivial relation between binary statistics and weights, which leads to important correlations between degrees and strengths in networks belonging to these ensembles~\cite{sagarra2013smm} which are also present in real data~\cite{Serrano2006}.

\section{Maximum likelyhood distributions}\label{app_2}

The probability distributions derived in this paper for networks belonging to the different described ensembles fulfil the maximum likelihood principle for networks~\cite{Garlaschelli2008a}. Indeed, the constraint point equations in \eqref{eq_saddle_general} can be understood as the equations resulting from maximizing the likelihood of the inverse problem of finding the set of values for the Lagrange multipliers $(\theta_T,\{\theta_q\})$ that maximize the likelihood of the observed adjacency matrix $\p{\hat{T}}$ to belong to each of the described models. In other words, defining the loglikelihood function of a network by
\begin{align*}
\mathcal{L}(\theta_T,\{\theta_q\}\vert \mathbf{\hat{T}}) = \ln \left(\prod_{ij} p_{ij}(\theta_T,\{\theta_q\}|\hat{t}_{ij} ) \right) = \sum \ln p_{ij}(\theta_T,\{\theta_q\}\vert \hat{t}_{ij})
\end{align*}
and maximizing this expression with respect to $(\theta_T,\{\theta_q\})$ one is lead to the equation \eqref{eq_saddle_general}. Explicitly,

\begin{widetext}
\begin{align}
\begin{split}
\partial_{\theta_q} \mathcal{L} &= \sum_{ij}  \left( c^{ij}_q \hat{t}_{ij} + \tilde{c}^{ij}_{q} \Theta(\hat{t}_{ij})- c^{ij}_q \ave{t_{ij}} - \tilde{c}^{ij}_{q} \ave{\Theta(\hat{t}_{ij})}\right) = \hat{C}_q - \ave{C_q}(\{\theta_q\}) \equiv  \Delta C_q\\
\partial_{\theta_q,\theta_q'} \mathcal{L} &= -\sum_{ij} \left( c^{ij}_q c^{ij}_{q'}\sigma^2_{t_{ij}}  + \tilde{c}^{ij}_q \tilde{c}^{ij}_{q'} \sigma^2_{\Theta(t_{ij})} \right) = 
-\sigma^2_{C_qC_{q'}} (\{\theta_q\})
\end{split}
\end{align}
\end{widetext}

which at the critical points lead to $\Delta C_q = 0\, \forall q$ and the condition of maximum with respect to all the variables is fulfilled (we also note that the problem in this form is concave). We thus see that the initial statement of the paper is confirmed: It is not enough to specify the constraints to fully define a maximum entropy ensemble, but one needs also to state the nature of its elements, since any maximum entropy ensemble will lead to a maximum likelihood description of a dataset.
There is not a "correct" ensemble to fix a given constraint, but the one that better describes the system that is being represented.

\section{Ensemble fluctuations}\label{app_3}

If the maximum entropy description provided here is to be successful, then the fluctuations of the obtained networks need to be bounded and have well defined statistics. In particular, we require that the associated entropy of the distribution and the statistics of the constraints possess finite first and second moments, and that their relative fluctuations around the average values need to be small in the limit of large sampling $\hat{T}$. Explicitly, we have,
\begin{align}
\begin{split}
\frac{\sigma^2_{C_q}}{\ave{C_q}^2} &= \frac{\sum_{ij} (c^{ij}_q)^2 \ave{t_{ij}}}{\left( \sum_{ij} c^{ij}_q \ave{t_{ij}} \right)^2} + \dfrac{a}{M}\frac{\sum_{ij} (c_q^{ij})^2 \ave{t_{ij}}^2 }{\left ( \sum_{ij} c_q^{ij} \ave{t_{ij}}\right )^2} \\
&= \frac{\sum_{ij} (c^{ij}_q)^2 \ave{t_{ij}}}{\left( \sum_{ij} c^{ij}_q \ave{t_{ij}} \right)^2} + \dfrac{a}{M} \frac{1}{1+ \alpha_q}\\
\alpha_q &\equiv \frac{\sum_{ij,kl |k\neq i, l\neq j}  c_q^{ij} c_q^{kl}\ave{t_{ij}}\ave{t_{kl}}}{\sum_{ij} (c_q^{ij})^2 \ave{t_{ij}}^2}.
\end{split}
\end{align}
where $a=0$ for ME case, $a=1$ for $W$ case and $a=-1$ for $B$ case.
We thus see that the fluctuations only disappear for large sampling in the linear case given by equation \eqref{eq_fij_linear} for the ME description. By construction, the constraints are extensive in the occupation numbers $t_{ij}$, thus when the number of events $T$ grows their average value $\ave{C_q}$ must also grow~\cite{sagarra2013smm}, yet only for the ME case we have $\ave{t_{ij}} \propto \hat{T}$ and thus only in this case relative fluctuations decay as $\hat{T}^{-1}$. Otherwise, the maximally random allocation of events will be made as homogeneous as possible among the states while preserving the constraints, hence $\alpha_q$ will in general be a large number (the denominator in the sum has $L$ terms while the numerator has $L(L-1)$, being $L$ the number of available node pairs for the allocation) and relative fluctuations will be bounded and $\mathcal{O}(M^ {-1})$. For similar reasons, one expects the first term to vanish for large sampling. For very large number of layers, then the ensembles become equivalent to the ME case, and fluctuations vanish in the large sampling limit~\cite{sagarra2014efn}. 

For the case where any binary constraint is additionally imposed (Appendix \ref{app_1}), the relative fluctuations of the binary structure dominate the statistics in the large sampling limit, and despite being bounded, these never vanish~\cite{sagarra2013smm}.

Concerning the associated Gibbs-Shannon entropy of the ensembles, since the occupation number statistics are independent, we have the random variable $\ln P(\mathbf{T})= \ln \prod_{ij} p_{ij}(t_{ij}) = \sum \ln p_{ij}(t_{ij})$ (associated to a given network instance) is a sum of independent contributions which in all cases studied (Poisson, Negative Binomial and Binomial) have well defined first and second moments when averaged over the ensemble. Hence, the statistics of $\ln P(\mathbf{T})$ will be Gaussian, and no extreme outliers are expected. This indicates that the total average number of possible network instances compatible with a given set of constraints is a well defined quantities, and one can define a {\em typical} network structure representing the ensemble (unlike other studied cases in the literature~\cite{bianconi2014a}).

\section{Calculation of degeneracy terms}\label{app_4}
\subsection{Layer degeneracy:}
For each state $ij$ out of the possible $L(N)=N^2$ node-pairs ($N(N-1)$ if not accepting self-loops) one needs to consider the process of allocating $t_{ij}$ events in $M$ possible distinguishable levels. For the W case this corresponds to the \emph{urn problem} of placing $t_{ij}$ identical balls in $M$ distinguishable urns. For the B case one faces the problem of selecting groups of $t_{ij}\leq M$ urns out of a set of $M$ urns and finally for the ME case one must count how to place $t_{ij}$ distinguishable balls in $M$ distinguishable urns. These problems are well known and their solution leads to the second column in \ref{table_deg_terms}, with the product over $ij$ representing the fact that the allocation among the layers for each node-pair is independent.

\subsection{Event degeneracy:}
For this calculation one only needs to take into account the distinguishable case (otherwise there is no degeneracy). Such a case, however is controversial to analyze. The correct counting of configurations in a Grand-Canonical ensemble is an issue spanning more than a century (see \cite{Swendsen2015} and references therein for details and extended discussion), ever since Gibbs used it to establish the relation in classical statistical mechanics between the Canonical and Grand-Canonical Ensembles of an ideal gas. 


Grand Canonical ensembles of networks can be faced in many ways. The usual view is to imagine a collection of $\mathcal{N}$ copies of a system in where to distribute $F$ events in such a way that there are $\hat{T} = F/\mathcal{N}$ events on average in each copy~\cite{Pathria1996}. In this framework, the probability to obtain a particular copy with $T=\sum_{ij} t_{ij}$ events and a set of constraints $\{C_q(\mathbf{T})\}$ reads,
\begin{equation}\label{eq:p_T}
P(\mathbf{T}) \propto e^{\theta_T T} e^{\sum_{q} \theta_q C_q(\mathbf{T})}.
\end{equation}
The prior expression is however related to the probability to obtain a \emph{given configuration} of $\mathbf{T}$, regardless whether 
it is unique or not (several configurations can give rise to the same $\mathbf{T}$). For the case of distinguishable events, there are $\binom{F}{T}$ different ways of obtaining the same number of events $T$ among the set of copies and $T!/\prod_{ij}t_{ij}!$ ways of distributing them to obtain a given adjacency matrix $\mathbf{T}$, hence one must consider an additional term in expression \eqref{eq:p_T},
\begin{equation}
\mathcal{D}(\mathbf{T})_{Events} = \binom{F}{T} \frac{T!}{\prod_{ij} t_{ij}!} .
\end{equation}
For the case with linear constraints of the form in equation~\eqref{eq_fij_linear}, the system partition function $\mathcal{Z}$ reads ($z_{ij} \equiv e^\theta_T \prod_{ij} e^{\theta_q c_q^{ij}}$),
\begin{widetext}
\begin{equation}
\begin{split}
\mathcal{Z} &= \sum_{\{\mathbf{T}\}} \mathcal{D}(\mathbf{T}) \prod_{ij}z_{ij}^{t_{ij}} = \sum_{T=0}^F \binom{F}{T} \sum_{\{\mathbf{T}| \sum_{ij} t_{ij}=\hat{T}\}} \frac{T!}{\prod_{ij} t_{ij}!} \prod_{ij} (Mz_{ij})^{t_{ij}} =  \sum_{T=0}^F \binom{F}{T} (M \sum_{ij} z_{ij})^T = (1+ M \sum_{ij} z_{ij})^F.
\end{split}
\end{equation}
If we add the strong condition that the ensemble average number of events has to be equal to $\hat{T}$, a scaling of $M\sum_{ij} z_{ij}$ on the total number of events $F$ distributed among the copies $\mathcal{N}$ is made apparent,
\begin{equation}
\begin{split}
\ave{T} &= \partial_{\theta_T} \ln \mathcal{Z} = \sum_{ij} \ave{t_{ij}} = F \frac{M \sum_{ij}z_{ij}}{1+M \sum_{ij}z_{ij}} = \hat{T} \implies M\sum_{ij}z_{ij} = \frac{\hat{T}/F}{1- \hat{T}/F}.
\end{split}
\end{equation}
Wrapping together the previous expressions and considering that the number of copies is arbitrary, one can imagine the limit where it goes to infinity keeping $\hat{T}$ constant. This amounts to consider $F$ infinitely large too,
\begin{equation}
\mathcal{Z} = \lim_{F \to \infty} \left(1 + \frac{\hat{T}/F}{1-\hat{T}/F} \right)^F = e^{\hat{T}} = \prod_{ij} e^{\ave{t_{ij}}} = \prod_{ij} \mathcal{Z}_{ij},
\end{equation}
which leads to a factorizable partition function of the form in equation~\eqref{eq_GC_ind} which does not depend on the number of copies of the system, as neither does its associated probability,
\begin{equation}
\begin{split}
P(\mathbf{T}) &= \lim_{F\to \infty} \frac{\mathcal{D}(\mathbf{T}) \prod_{ij}z_{ij}^{t_{ij}}}{\mathcal{Z}} = \lim_{F\to\infty} \frac{\binom{F}{T}\frac{T!}{\prod_{ij}t_{ij}!} \prod_{ij} \left(\frac{\ave{t_{ij}}}{(F-\hat{T})} \right)^{t_{ij}}}{\left(1 - \frac{\hat{T}}{F} \right)^{-F}} = \prod_{ij} \frac{\ave{t_{ij}}^{t_{ij}}}{t_{ij}!} e^{-T} \lim_{F\to \infty} \left(\frac{F-\hat{T}}{F-T} \right)^{F-T} = \\&=\prod_{ij} \frac{\ave{t_{ij}}^{t_{ij}}}{t_{ij}!} e^{-T} e^{-\hat{T} + T}=\prod_{ij} p_{ij}^{ME}(t_{ij}).
\end{split}
\end{equation}
\end{widetext}
We have thus reached the same independent Poisson probabilities as the ones obtained by taking an \emph{effective} factorizable degeneracy term $\hat{T}! \prod_{ij} (M^{t_{ij}}/t_{ij}!)$ (see equation \eqref{eq_all_stats}).

These results are in accordance with previous works~\cite{sagarra2013smm} where the complete equivalence between Canonical (ensembles with soft linear constraints as in equation~\eqref{eq_fij_linear} but with $T=\hat{T}$ fixed for every network in the ensemble) and Micro-canonical ensembles (ensembles where all constraints are exactly fulfilled) of Multi-Edge networks was proven. The equivalence between Micro-canonical ensembles and Grand-canonical ensembles in Poisson form has also been validated by simulations for the case where strengths are fixed~\cite{sagarra2014efn}.

For non linear cases (such as the case with binary constraints or the B case with distinguishable events), the \emph{effective} degeneracy term is an approximation, since the complete calculation using partial sums where $T$ is exactly fixed cannot be performed. Approximating $T!$ by $\hat{T}!$ amounts to consider that the possible fluctuations of the macroscopic variable $T$ are caused by the state independent fluctuations of the microscopic structure given by $\{t_{ij}\}$. This however leads to the same statistics emerging from the leading order terms in $\hat{T}$ of the system partition function computed using a Micro-canonical formalism (see~\cite{sagarra2013smm}).
%
%

\bibliography{referencessub}

\newpage

\begin{widetext}

\begin{center}
\textbf{\large Role of adjacency matrix degeneracy in maximum-entropy-weighted network models: Supplementary Material}
\end{center}

\section{Datasets}
\subsection{Taxi data}
The taxi data are the same as those used in other studies \cite{Santi14,sagarra2015a}, which represents the aggregated number of taxi trips between road intersections performed within the borough of Manhattan for the year 2012. Each node is an intersection ($N=4090$) and $\hat{T} = \sum_{ij} \hat{t}_{ij} \simeq 150 \cdot 10^6$ trips are generated in this interval. The network contains a negligible fraction of self-loops and is directed. For the aggregation of multiplexes, we have considered that the network contains $M=365$ different daily temporal snapshots.
\subsection{WTN}
The data for the World Trade Network have been obtained from \cite{Barbieri2009}. We consider the aggregated snapshot for the trade (counted in millions of dollars) for the year 2002 between $N=184$ countries appearing in the dataset, corresponding to $M=100$ different trade commodities considered as layers. Since the dataset is incomplete, we have proceeded as follows for each pair of countries: If a flow is missing in both directions, we ignore the edge $\hat{t}_{ij} = \hat{t}_{ji} = 0$, furthermore, if only one direction is missing, the non-missing value is copied for both directions. Finally, a threshold of $\hat{t}_{min} = 1$ is applied and all values are truncated to the nearest integer to enforce the condition $\hat{t}_{ij} \in \mathbb{N} \, \forall i,j$. The network is directed and does not contain self loops nor nodes for which the incoming and outgoing degrees are zero.
\subsection{Multiplex European Airlines}
The data for the aggregated multiplex of flight connections between European cities have been obtained from \cite{Cardillo2013}. It has $N=417$ nodes which represent airports and weights represent the existence of a connection by a given airline between two airports. The layers are thus the $M=37$ different airlines present in the dataset, and the adjacency matrix has been obtained by aggregating all the binary layers, $\hat{t}_{ij} = \sum_m \Theta(\hat{t}_{ij}^m)$. The network is undirected but is represented as directed ($\hat{t}_{ij} = \hat{t}_{ji}\,\forall i,j$) and does not contain self loops nor nodes for which the degrees are zero.

The strength distribution for all the datasets is shown in figure \ref{fig:s_distro} and network quantities are summarized in table \ref{table_net_details}.

\begin{table}[htbp]
\begin{center}
\begin{tabular}{l | c c c c c c c c} 
Dataset & N & $\overline{\hat{s}}$ & $E/L = \sum_{j} \hat{k}_j^{out}/L$ & $T/L$ &$\frac{\sigma_{\hat{t^+}}}{\overline{\hat{t^+}}}$ &$\frac{\sigma_{\hat{s}_{out}}}{\overline{\hat{s}_{out}}}$ & $\frac{\sigma_{\hat{k}_{out}}}{\overline{\hat{k}_{out}}}$&Self-loops\\\hline
Taxis & 4090 & 35569.357 & 0.433 &8.697& $1.0$ & $1.5$ &$0.7$ & Yes\\
WTN & 184 & 34350.603 & 0.323 &187.708& $10.0$ & $2.8$ &$0.8$ & No\\
Airlines & 417 & 17.206 & 0.034 &0.041& $1.3$ & $1.6$ &$1.5$ & No
\end{tabular}
\caption{Network details on the used datasets. The symbol $\bar{x}=N^{-1} \sum_i x_i$ indicates the graph-average of quantity $x$, $L=N(N-1)$ for the self-loops allowed case and $L=N^2$ otherwise, see section \ref{sec:mags} for details on each quantity.}
\label{table_net_details}
\end{center}
\end{table}

\begin{figure}[htbp]
\begin{center}
\includegraphics[width=0.8\textwidth]{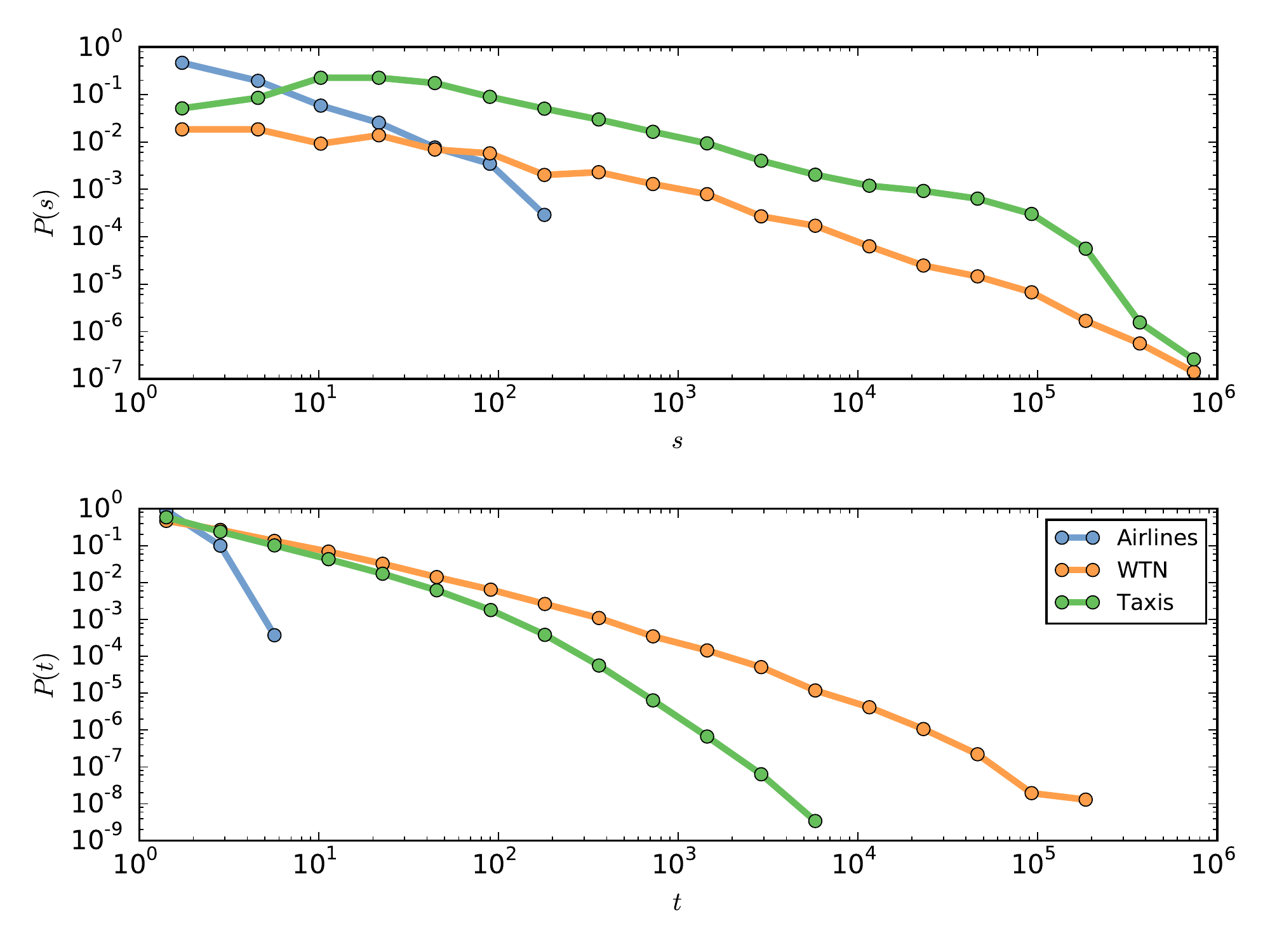}
\caption{(color online) Outgoing strength and existing weight distribution for the considered datasets smoothed with log binning. All distributions are fat-tailed except the existing weight for the Airline dataset, which is bounded by the value $M=37$ by construction.}
\label{fig:s_distro}
\end{center}
\end{figure}

\section{Measured quantities}\label{sec:mags}
In the paper, we compute the differences between models for different first and second order rescaled network metrics, detailed in the following (we only present the outgoing version of each quantity):
\begin{itemize}
\item Scaled binary degrees: Represents the percentage of the network connected by at least a single event to a given node.
\begin{align*}
\frac{k_i}{N} = \dfrac{1}{N}\sum_j \Theta(t_{ij}).
\end{align*}
\item Non-zero weight distribution: Distribution of occupation numbers per existing links.
\begin{align*}
P(t) = \frac{\sum_{ij} \delta_{t,t_{ij}}}{\sum_{ij} \Theta(t_{ij})}.
\end{align*}
\item Scaled graph-average weight and graph-average binary connection probability of edges as a function of product of incoming and outgoing strengths: We use this metric to analyze the correlations between the binary connection probability and the average weight of links, as a function of the "importance" of each link as measured by the product of the nodes' strengths.
\begin{align*}
\begin{split}
\frac{\bar{t}}{\hat{T}}(\hat{s}^{out} \hat{s'}^{in}) &= \frac{1}{\hat{T} N_s N_{s'}} \sum_{ij|\hat{s}^ {out}_i=\hat{s},\hat{s}^ {in}_j=\hat{s'}} t_{ij} \\
\bar{\Theta(t)}(\hat{s}^{out} \hat{s'}^{in}) &= \frac{1}{N_s N_{s'}} \sum_{ij|\hat{s}^ {out}_i=\hat{s},\hat{s}^ {in}_j=\hat{s'}} \Theta(t_{ij}).
\end{split}
\end{align*}
\item Disparity: This bounded quantity $Y_{2,i} \in (k_i^{-1}(s_i),1]$ indicates the homogeneity in the distribution of events among the edges emerging from a given node. The higher this value, the more concentrated the events are on a predominant link.
\begin{align*}
Y_{2,i} = \frac{\sum_j t_{ij}^2}{(\sum_j t_{ij})^2.}
\end{align*}
\item Scaled average neighbor weighted strength: This quantity indicates the average strength of the neighbors of a nodes, weighted by the events of each connection. In this version, we have rescaled the quantity by the uncorrelated expectation $\left. s^{w}_{nn,i} \right|_{uncorr} = \frac{\overline{\hat{s}^2}}{\overline{\hat{s}}} = \frac{\sum_{i} (\hat{s}^{out}_i)^2}{\hat{T}}$.
\begin{align*}
s^w_{nn,i}\frac{\overline{\hat{s}}}{\overline{\hat{s}^2}} = \frac{\overline{\hat{s}}}{\overline{\hat{s}^2}} \frac{\sum_j t_{ij} s^{in}_j}{ s^{out}_i}
\end{align*}
\item Scaled average neighbor degree: These quantities represents node degree correlations, at the binary connection level.
\begin{align*}
\frac{k_{nn,i}}{N} = \frac{\sum_j \Theta(t_{ij}) k^{in}_j}{N k^{out}_i}
\end{align*}
\item Sorensen Common part of Commuters index: This indicator is used to compute similarity between weighted matrices \cite{Lenormand2012} and we use the version (less prone to be affected by sampling) appearing in \cite{sagarra2015a}.
\begin{align*}
CPC_{model} = 2\frac{\sum_{ij} \min(\hat{t}_{ij},\ave{t_{ij}}^{model})}{\sum_{ij} \hat{t}_{ij} + \sum_{ij} \ave{t_{ij}}^{model}}.
\end{align*}
\end{itemize}
Some of these cases have the issue of low-strength nodes which do not appear in every simulation ($s=0$ or $k=0$), leading to an undefined value of such quantities. Hence to average over an ensemble for this cases, we only compute the conditioned mean on existing cases.

\section{Additional comparison between models and data}
We provide in this section additional details in the comparison between models and data. Table \ref{table_models} presents the graph-average error $\overline{\varepsilon_x} = N^{-1}\sum_i (\hat{x}_i - \ave{x_i})$ and graph-average standard deviation $\sigma_{\varepsilon_x} = \left(N^{-1}\sum_i (\hat{x}_i - \ave{x_i})^2\right)^{1/2}$ for different quantities displayed in the main text. Figure \ref{fig_more_nets} also shows the figures corresponding to the Airlines dataset.

The model values have been computed averaging over $r=10^4$ realizations for the WTN and Airline cases and $r=500$ for the Taxis respectively.

\begin{table}[tbp]
\begin{center}
\begin{tabular}{l |l |c c c c | c}
Dataset & Case & 
$\overline{\varepsilon_{k}}\pm\sigma_{k}$ & $\overline{\varepsilon_{Y_2}}\pm\sigma_{Y_2}$ &
$\overline{\varepsilon_{s^w_{nn}}}\pm\sigma_{s^w_{nn}}$ &
$\overline{\varepsilon_{k_{nn}}}\pm\sigma_{k_{nn}}$&
$CPC$
\\\hline
\multirow{4}{*} 
{Airlines} &
ME & $-0.23 \pm 2.72$&$ -0.01 \pm 0.11$&$ -1.93 \pm 27.96$&$-3.64 \pm 19.59$&0.329\\
&\textbf{B} ($M=37$) & $-0.18 \pm 2.65$&$-0.01 \pm 0.11$&$-1.60 \pm 27.94$&$-3.13 \pm 19.58$&0.329\\
&W ($M=37$) & $-0.28 \pm 2.80$&$-0.01 \pm 0.11$&$-2.27 \pm 28.00$&$-4.14 \pm 19.62$&0.329\\
&W  ($M=1$) & $-1.64 \pm 5.75$&$-0.01 \pm 0.11$&$-9.26 \pm 28.82$&$-14.42 \pm 20.18$&0.323 \\  \hline
\multirow{4}{*} 
{Taxis} 		&
\textbf{ME}  & $250.91 \pm 201.88$&$-0.00 \pm 0.01$&$26782 \pm 32260$&$491 \pm 422$&0.649\\
&B ($M=365$) & $159.72 \pm 192.03$&$0.00 \pm 0.01$&$84231 \pm 80803$&$465 \pm 474$&0.639\\
&W ($M=365$) & $283.80 \pm 216.95$&$-0.00 \pm 0.01$&$14147 \pm 31352$&$495 \pm 403$&0.643\\
&W ($M=1$) & $137.19 \pm 255.48$&$-0.00 \pm 0.01$&$-14341 \pm 27504$&$134 \pm 289$&0.624  \\\hline
\multirow{4}{*} 
{WTN} &
\textbf{ME} & $42.24 \pm 29.42$&$-0.12 \pm 0.18$&$59\cdot 10^3 \pm 202 \cdot 10^3$&$31.86 \pm 17.55$&0.62\\
&B ($M=100$) & -- & -- & -- & -- \\
&W ($M=100$) &$70.47 \pm 45.54$&$ -0.17 \pm 0.19$&$-194\cdot 10^3 \pm 224\cdot 10^3$&$45.28 \pm 23.50$&0.53\\
&W  ($M=1$)&$63.48 \pm 45.16$&$-0.16 \pm 0.19$&$-212\cdot 10^3\pm 227\cdot 10^3$&$35.08 \pm 24.09$&0.53 \\  \hline
\end{tabular}
\caption{Graph average differences between model and data quantities. Bold letters indicate the best performing model as indicated by the CPC index.}
\label{table_models}
\end{center}
\end{table}

\begin{figure}[tbp]
\begin{center}
\includegraphics[width=0.8\textwidth]{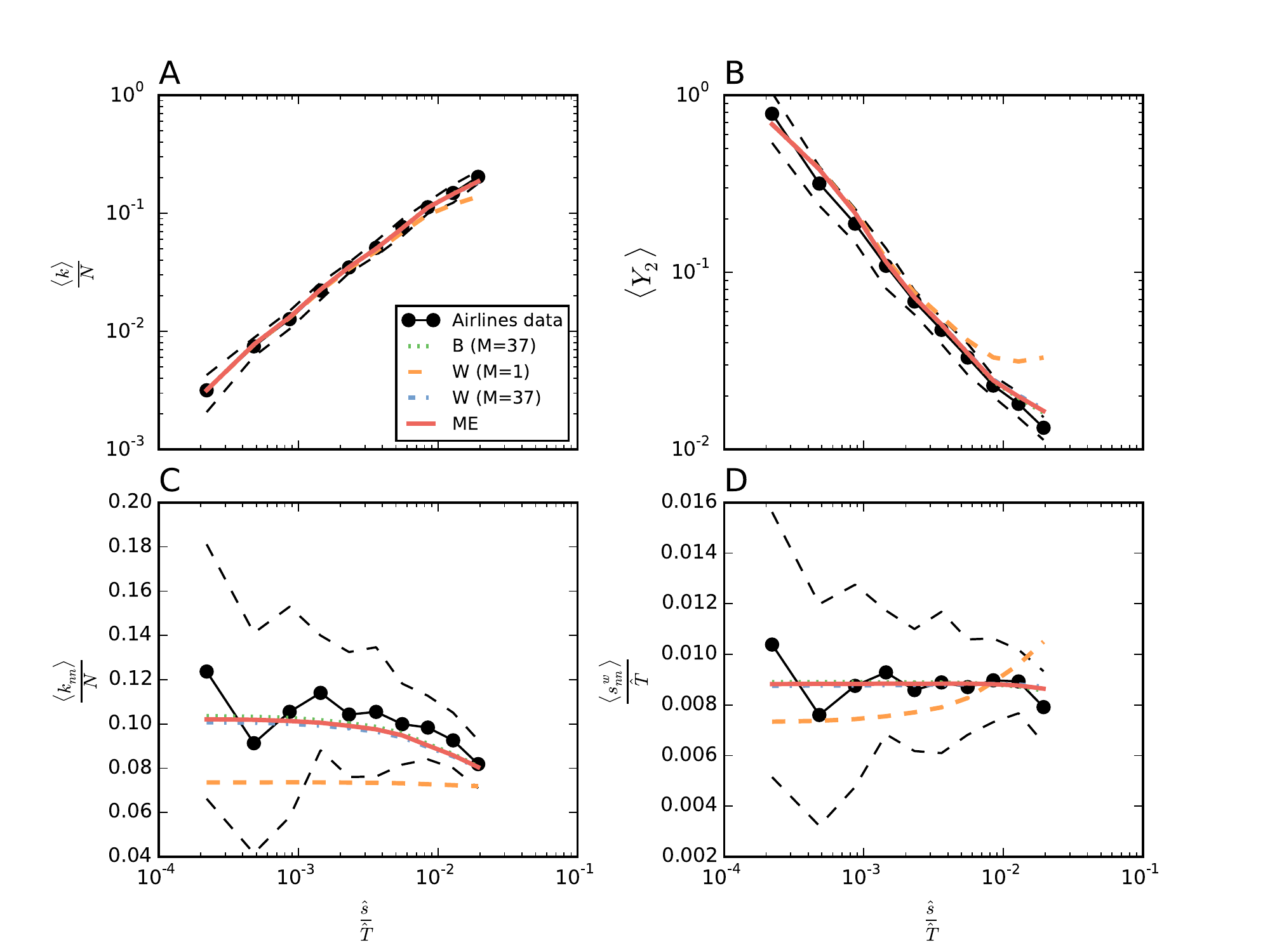}
\includegraphics[width=0.8\textwidth]{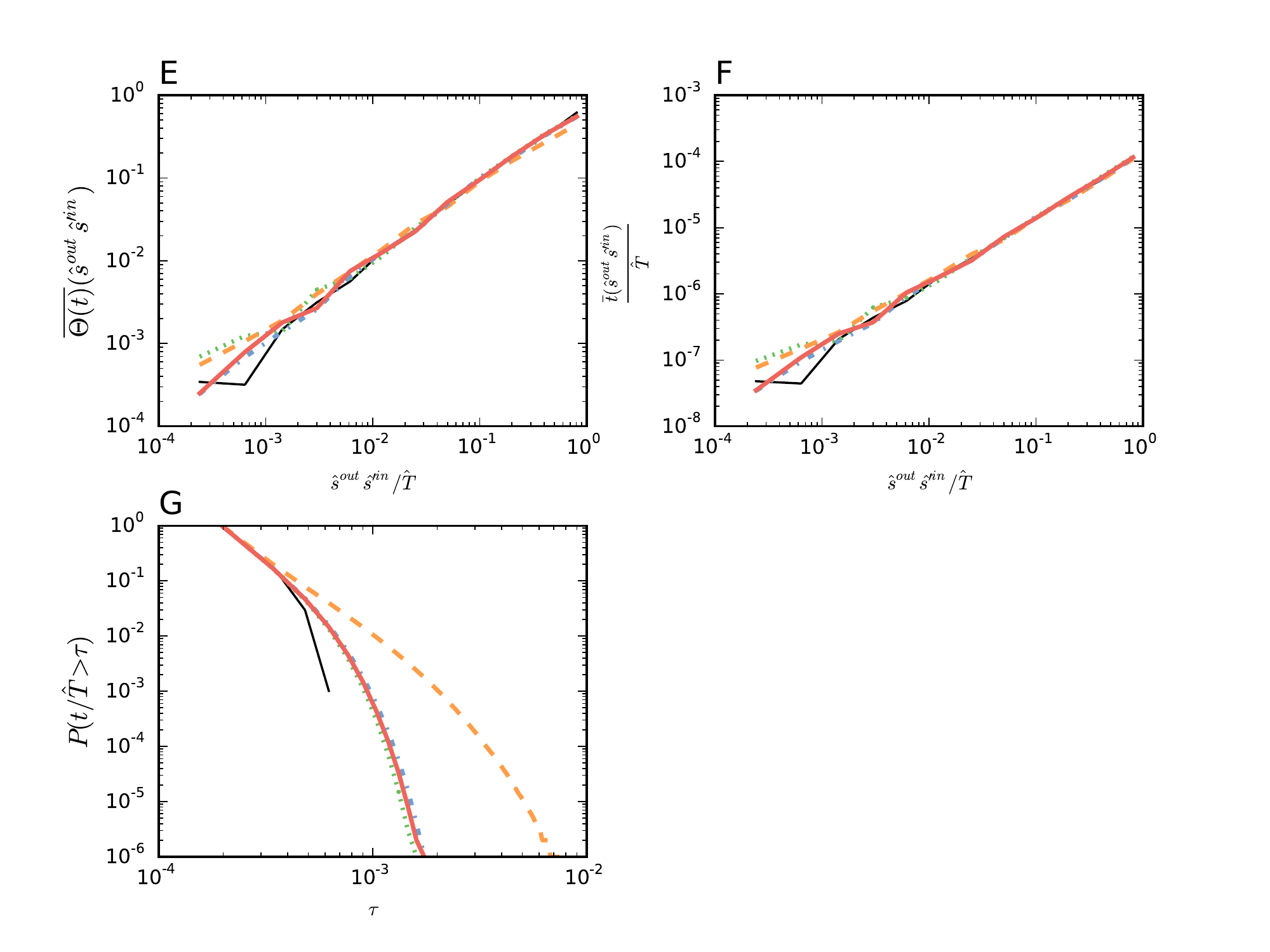}
\caption{Node and edge related metrics for the Airlines dataset over $r=10^4$ realizations, see section \ref{sec:mags} for details on the measured quantities. In this case, due to poor sampling, differences between ensembles are not clearly appreciable, yet the W case is clearly distinct from the others and from the real data, being also the B case the one closer to the empirical data.}
\label{fig_more_nets}
\end{center}
\end{figure}

\section{Solving the saddle point equations}
%

The only caveat related with the maximum entropy ensembles of networks is the fact that one needs to solve the saddle point equations related with each particular ensemble one has chosen.

Recovering the expressions in Appendix 2 of the main text,
\begin{align}\label{eq_loglik}
\begin{split}
\partial_{\theta_q} \mathcal{L} &= \sum_{ij} c^{ij}_q \left( \hat{t}_{ij} - \ave{t_{ij}}\right)  \equiv \Delta C_q\\
\partial_{\theta_q,\theta_q'} \mathcal{L} &=  -\sum_{ij} c^{ij}_{q'}c^{ij}_q \sigma^2_{t_{ij}}=- \sigma^2_{C_qC_{q'}}
\end{split}
\end{align}
we see that the concavity of the function to solve is assured in all the cases, yet the method to solve each system of equations will depend on the specifics of the problem at hand. In the general case, the difficulty of the problem depends on the number of constraints and varies with the particular restrictions of the considered case.

For the case of fixed strengths, one can approach this problem by maximizing the likelihood associated to every model, compared to the actual strength list provided $\{\hat{s}^{out},\hat{s}^{in}\}$. Hence, we need to solve the $N$ associated $\{x,y\}$ values considering $z_{ij} = x_i y_j$. So we have,
\begin{align}\label{eq_vars}
\begin{split}
c^{ij}_{q_{out}} = \delta_{q_{out}i} \qquad c^{ij}_{q_{in}} = \delta_{jq_{in}} \qquad e^\theta_{q_{out}} \equiv x_q \qquad e^\theta_{q_{in}} \equiv y_q.
\end{split}
\end{align}

In the following, we provide the details on the computational methods used to solve the saddle point equations, which for the particular case of fixed strengths discussed in the paper, is implemented in the freely available, open source package ODME \cite{ODME2014}.

\subsection{Multi Edge case}
This scenario is fully analytical \cite{sagarra2014efn} for the case where self-loops are accepted. Otherwise, the statistic is Poisson so $\ave{t_{ij}} = \sigma^2_{t_{ij}} = x_iy_j$ and one has the only restriction that $x_i\geq 0, y_i \geq 0 \forall i$,
\begin{align}
\begin{split}
\partial_{x_q} \mathcal{L}^{ME}_s =  \frac{\hat{s}^{out}}{x_q} - M\sum_j y_j  & \qquad \partial_{y_q} \mathcal{L}^{ME}_s =   \frac{\hat{s}^{in}}{y_q} - M\sum_i x_i  \\
\partial_{x_q x_l} \mathcal{L}^{ME}_s = -\delta_{ql}\frac{\hat{s}_{out}}{x_q^2} & \qquad \partial_{y_q y_l} \mathcal{L}^{ME}_s = -\delta_{ql} \frac{\hat{s}_{in}}{y_l^2}\\
\partial_{y_q x_l} \mathcal{L}^{ME}_s &= -M.
\end{split}
\end{align}
The saddle point equation can be then safely solved using algorithm \ref{alg:balancing} with
\begin{align}
F(\{x,y\},\{\hat{s}^{out},\hat{s}^{in}\})=
\left \{ \begin{array}{l } 
x_q^{n+1} = \frac{\hat{s}^{out}_q}{M \sum_j y_j^{n}} \\
y_q^{n+1} = \frac{\hat{s}^{in}_q }{M \sum_i x_i^{n+1}} 
\end{array} \right.
\end{align}

\begin{algorithm}[htbp]
 \KwData{Constraints list $\{\hat{C}_q\}$, tolerance on constraints $\varepsilon_Q$, tolerance on variables $\varepsilon_z$ and initial guess $\{z^{ini}\}$.}
 \KwResult{Set of Lagrange multipliers $\{z_q\}$}
 Set $z_q^0=z^{ini}_q$, $z_q^1=0$ $\forall q \in [1...Q]$. Set $n=0$.\;
 \While{$\max |\hat{C}_q - \ave{C_q}(\{z_q^{n}\})|>\varepsilon_Q$ or $\max |z_q^{n+1}-z_q^{n}|>\varepsilon_z$}{
  	\qquad$z_q^{n+1} = F_q(\{z^{n}\},\{\hat{C}_q\})$\;
  	\qquad$n = n+1$\
 }
 \caption{Balancing algorithm to solve saddle point equations for ME and B cases.}
 \label{alg:balancing}
\end{algorithm}

\subsection{Binary case}
In this case the restriction is analogous to the previous one and we have,
\begin{align}
\begin{split}
\partial_{x_q} \mathcal{L}^{B}_s =   \frac{\hat{s}^{out}}{x_q} - M\sum_j \frac{y_j}{1+x_qy_j}  & \qquad \partial_{y_q} \mathcal{L}^{B}_s =   \frac{\hat{s}^{in}}{y_j} - M\sum_i \frac{x_i}{1+x_iy_q}  \\
\partial_{x_q x_l} \mathcal{L}^{B}_s = -\frac{\delta_{ql}}{x_q^2} \left( \hat{s}_q^{out} - M\sum_j \left(\frac{ x_qy_j}{1+x_qy_j}\right)^2 \right)& \qquad \partial_{y_q y_l} \mathcal{L}^{ME}_s = - \frac{\delta_{ql}}{y_q^2} \left( \hat{s}_q^{in} - M\sum_i \left(\frac{ x_iy_q}{1+x_iy_q}\right)^2 \right)\\
\partial_{y_q x_l} \mathcal{L}^{B}_s &= -\delta_{ql} \frac{M}{(1+x_qy_l)^2}
\end{split}
\end{align}
Although the problem cannot be shown to be strictly concave in this form, the saddle point equations can again be solved using algorithm \ref{alg:balancing} with the relation
\begin{align}
F(\{x,y\},\{\hat{s}^{out},\hat{s}^{in}\})=
\left \{ \begin{array}{l } 
x_i^{n+1} = \frac{\hat{s}^{out}_i}{ \sum_j \frac{y_j^{n}}{1+x_i^{n}y_j^{n}}} \\
y_j^{n+1} = \frac{\hat{s}^{in}_j}{  \sum_i \frac{x_i^{n+1}}{1+x_i^{n+1}y_j^{n}}}.
\end{array} \right. 
.
\end{align}

Some convergence problems can be encountered using the algorithm \ref{alg:balancing} if the larger values of the strength approach the limit $\max \{ \hat{s}^{out},\hat{s}^{in}\} = M$, but in this case a simple, positively bounded, unconstrained gradient descent method has been implemented to solve the problem. 

\subsection{Weighted case}
The weighted case is considerably more complicated than the previous ones, since it includes the restriction that $0 \leq x_iy_j  < 1\, \forall \, i,j$ and hence the maximization is performed on a non-convex domain. The balancing approach (algorithm \ref{alg:balancing}) is then not satisfactory, since there is no explicit enforcement for the values $\{x,y\}$ to remain in the domain of the Loglikelihood function one wants to maximize. The scalar function being considered is
\begin{align}
\mathcal{L}^W_s  = K(M,\{\hat{t}_{ij}\}) + M \sum_{ij} \ln (1-x_i y_j) +& \sum_i \hat{s}^{out}_i \ln x_i + \sum_j \hat{s}^{in}_j \ln y_j
\end{align}
with derivatives,
\begin{align}
\begin{split}
\partial_{x_q} \mathcal{L}^{W}_s =  \frac{\hat{s}^{out}}{x_q} - M\sum_j \frac{y_j}{1-x_qy_j} & \qquad \partial_{y_q} \mathcal{L}^{W}_s =  \frac{\hat{s}^{in}}{y_j} - M\sum_i \frac{x_i}{1-x_iy_q} \\
\partial_{x_q x_l} \mathcal{L}^{W}_s = -\frac{\delta_{ql}}{x_q^2} \left( \hat{s}_q^{out} + M\sum_j\left( \frac{x_qy_j}{1-x_qy_j}\right)^2 \right)& \qquad \partial_{y_q y_l} \mathcal{L}^{W}_s = - \frac{\delta_{ql}}{y_q^2} \left( \hat{s}_q^{in} + M\sum_i \left(\frac{x_iy_q}{1-x_iy_q}\right)^2 \right)\\
\partial_{y_q x_l} \mathcal{L}^{W}_s &= -\delta_{ql} \frac{M}{(1-x_qy_l)^2}
\end{split}
\end{align}
subject to the conditions that,
\begin{equation}\label{eq_cond}
 0 \leq x_iy_j\leq 1\, \forall i,j \in \{1...N\}.
\end{equation}
In principle, the problem is concave, and thus finding a solution to the saddle point equations gives the global maximum. Sadly, for real cases this concavity is lost as soon as the explicit domain constraint is included $0\leq x_iy_j < 1$. Hence, there is no general algorithm that can be applied with assured results, and obtaining a solution to the saddle point equations will not be guaranteed in all cases (specially for large $N$ or a very skewed distribution of strengths).

We thus deal with a large scale, non-concave (non-convex), bounded and constrained maximization (minimization) problem.

\subsubsection{Preconditioning}\label{sec_uconstr}
Basically, the difficulty of the problem derives from the form of the strength sequence $\{\hat{s}^{out},\hat{s}^{in}\}$. The more skewed this distribution is, the more difficult the problem is to solve, for a given fixed $N$. For \textit{easy} problems, a good way to pre-condition the problem is to first solve the easier, bounded, unconstrained problem of finding,
\begin{equation}\label{eq_uconstr_easy}
\begin{split}
&\min \left [ - \mathcal{L}^{W}_s(\mathbf{x})\right]\\
& 0 \leq x_i < \alpha \quad i \in [1...N] \\
& 0 \leq y_j < \alpha^{-1} \quad j \in [1...N] \\
& \alpha \in \mathbb{R}^+.
\end{split}
\end{equation}
The problem of this method is that it does not consider all the available phase space (see figure \ref{fig_phase}): The solution lies in the hyper-volume defined by the axis and $y_{max}(x_{max}) = x_{max}^{-1}$, which is a larger volume than that defined by the axis and $x_{max} \leq \alpha,y_{max}\leq \alpha^{-1}$. Usually, a good choice is $\alpha=1$. If the distribution of of strengths is very skewed, the optimal solution most likely lies outside the second area, but the suboptimal solution within this region serves as preconditioning for the complete maximization problem thanks to the convexity of the function (without considering the domain).

We have implemented this preconditioning procedure using a truncated Newton TNC method \cite{nocedal2006} from the Scipy suite \cite{scipy}.

\begin{figure}[htbp]
\begin{center}
\includegraphics[width=0.8\textwidth]{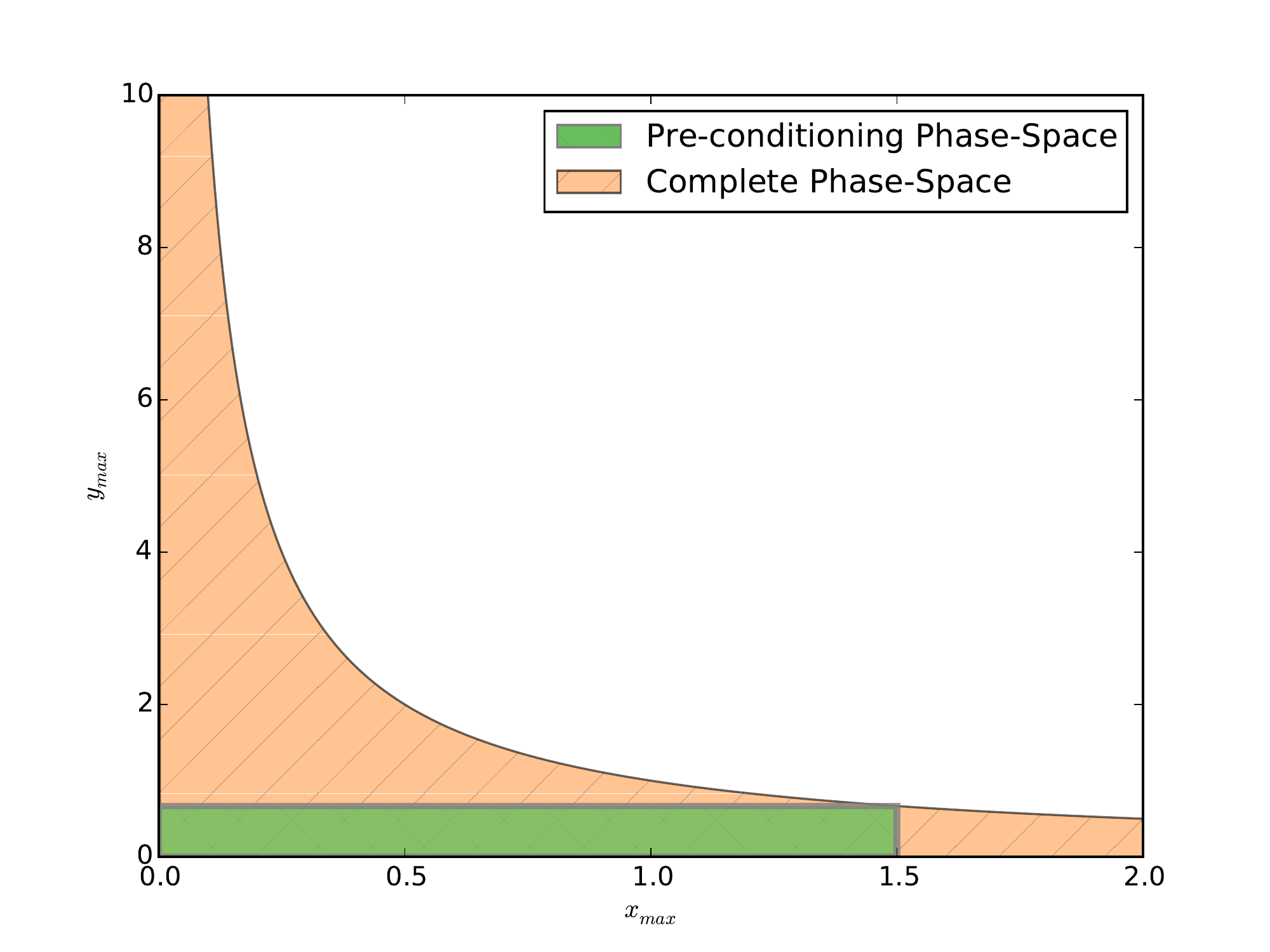}
\caption{Sketch of a plane projection of the phase space hyper volume with $\alpha=1.5$ for the maximization problem in the W case. The preconditioning method looks for solutions inside the area delimited by the green rectangle. The orange area represents the domain of the $\mathcal{L}^{W}_s$ function, which is clearly non-convex.}
\label{fig_phase}
\end{center}
\end{figure}

\subsubsection{Constrained problem}
Since the loglikelihood function is not defined outside of the domain, we use an interior point method to solve the problem adding $L$ non-linear inequalities of the form $0\leq x_i y_j < 1$ ($L=N^2$ for the case without selfloops and $L=N(N-1)$ otherwise). The implementation is done in CVXOPT \cite{CVXOPT}, but it has an obvious limitation given by the use of memory, which grows very fast with the number of nodes of the given network. Additionally, as early mentioned, the convergence of the algorithm is not assured due to the non-convexity of the complete problem, yet in our cases we obtained very satisfactory cases for the different cases analyzed.

\subsection{Precision}
For all the cases considered, we analyze the precision of our solving approach by computing the euclidean norm of the absolute error and the maximum of the absolute relative error among the nodes,
\begin{align}
|\Delta C | = \sqrt{\sum_q (\Delta C_q)^2} = \sqrt{\sum_q (\hat{C}_q - \ave{C}_q)^2} \qquad 
\varepsilon_{max} = \max \frac{| \Delta C_q |}{\hat{C}_q}.
\end{align}
The resulting values for each example are reported in table \ref{table_norms}.
\begin{table}[htbp]
\begin{center}
\begin{tabular}{l |l || l l l } 
Dataset & N & Case & $|\Delta C|$ & $\varepsilon_{max}$\\\hline
Airlines&\multirow{4}{*} 
{417} &
ME & $4 \cdot 10^{-12}$ & $10 \cdot 10^{-15}$\\
&&B ($M=37$) & $9\cdot 10^{-12}$ & $4\cdot10^{-14}$\\
&&W ($M=37$) & $1\cdot 10^{-9}$ & $1\cdot 10^{-12}$\\
&&W  ($M=1$) & $2\cdot 10^{-9}$ & $3\cdot 10^{-12}$ \\  \hline
Taxis&\multirow{4}{*} 
{4090} &
ME     & $1 \cdot 10^{-8}$ & $6 \cdot 10^{-14}$\\
&&B ($M=365$) &  $1\cdot 10^{-7}$ & $1 \cdot 10^{-13}$ \\
&&W ($M=365$) &  $7 \cdot 10^{-6}$ & $5 \cdot 10^{-12}$ \\
&&W ($M=1$) & $0.05$ & $4\cdot 10^{-8}$\\  \hline
WTN&\multirow{4}{*} 
{188} &
ME & $5 \cdot 10^{-10}$ & $2 \cdot 10^{-15}$\\
&&B ($M=100$) & -- & -- \\
&&W ($M=100$) & $3 \cdot 10^{-5}$ & $5\cdot 10^{-6}$\\
&&W  ($M=1$)& $7\cdot 10^{-4}$ & $1 \cdot 10^{-4}$\\  \hline

\end{tabular}
\caption{Results of the maximization problem. Norms associated to the best solution for each dataset and synthetic data. Missing values for Binary case indicate $s_{max}>MN$, hence the model is not applicable.}
\label{table_norms}
\end{center}
\end{table}

\section{Availability of code and Data}
The basic algorithms and codes used in this paper are all part from the freely available \textbf{Origin Destination Multi Edge} analysis package (ODME) \cite{ODME2014}.
The data for the WTN has been obtained from \cite{Barbieri2012b}, the airline network data from \cite{Cardillo2013} and the Taxi data has been obtained from the New York Taxi and Limousine Commission via Freedom of Information Law request.

\end{widetext}
\end{document}